\documentclass[twocolumn,nopacs,preprintnumbers,amsmath,amssymb]{revtex4}

\usepackage{graphicx}
\usepackage{dcolumn}
\usepackage{bm}

\begin{document}

\preprint{APS/123-QED}

\title{Undulation Amplitude of a Fluid Membrane \\
in a Near-Critical Binary Fluid Mixture \\
Calculated beyond the Gaussian Model \\ Supposing Weak Preferential Attraction}

\author{Youhei Fujitani}
 \email{youhei@appi.keio.ac.jp}
\affiliation{School of Fundamental Science and Technology,
Keio University, 
Yokohama 223-8522, Japan}

\date{\today}

\begin{abstract}
We calculate the mean square amplitude of the shape fluctuation -- an equal-time correlation -- of
an almost planar fluid membrane immersed in 
a near-critical binary fluid mixture.
One fluid component is usually preferentially attracted by the membrane, and
becomes more concentrated around it because of the near criticality.
This generates osmotic pressure, which influences the amplitude.  The amplitude is also affected by the 
reversible dynamics of
the mixture, which moves with the membrane.
By assuming the Gaussian free-energy functional and weak preferential attraction,
the author previously showed that
a new term is added to the restoring force of the membrane and 
tends to suppress the amplitude.  
Not assuming both of them, but still focusing on modes with wavelength longer than the 
correlation length, we here calculate the amplitude of a tensionless membrane.
First, within the Gaussian model, we 
solve the governing equations to 
show that, for long wavelength, the additional term becomes predominant,  
although decreased hydrodynamic effects make
its numerical factor approximately half that
of the previous result.  The change in the term turns out not to be monotonic with the wavelength, which is mainly caused by
the change in the induced mass. 
Second, assuming the critical composition far from the membrane,  we
calculate the amplitude beyond the regime of 
the Gaussian model.
The result coincides roughly with the
corresponding result in the Gaussian model  
if the correlation length is interpreted as one close to the membrane. 
\end{abstract}
\maketitle

\section{\label{sec:intro}Introduction}
A fluid membrane is a two-dimensional fluid with bending rigidity \cite{canh,helfbend},
and is usually composed of amphiphilic molecules.
A typical example is the lipid-bilayer membrane in 
biomembranes of cells \cite{sing}, where the polar head groups are arranged 
outwards and in contact with
aqueous environments.  The thermal undulation, or
shape fluctuation, of almost planar membranes at equilibrium can explain the flicker phenomenon of red blood cells \cite{broc}.
When the cell is not swollen, the membrane is freely suspended in the three-dimensional fluid and
the mean lateral tension of the membrane is negligible \cite{fritz,shmi,shiba}. 
Such a tensionless membrane, whose restoring force comes from only the bending rigidity,
becomes floppy enough 
to lose its orientation if the membrane is sufficiently large \cite{helf3,helf2,sorn}.
In a bilayer membrane composed of sodium dodecyl sulfate,
pentanol, and water, the nonpolar chains
are arranged outwards in some organic solvent \cite{bel,diat,lei2}.  
This membrane, as well as the lipid-bilayer membrane, 
has a thickness of $4-5$ nm. \\

\begin{figure}[b]
\includegraphics[width=8cm]{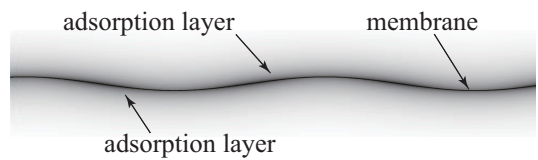}
\caption{\label{fig:sample} Schematically drawn side view of a fluctuating membrane (solid curve).
The gray scale reflects the deviation of the 
mass-density difference between the two components from its value far from the membrane, and the shaded region
represents the adsorption layer.  }
\end{figure}

Consider
a fluid membrane immersed in a near-critical binary fluid mixture in the homogeneous phase.
The membrane would preferentially attract
one component of the mixture, such as a wall \cite{Cahn,binder,bray,diehl86,burk,diehl97,JCP}.
The preferred component is more concentrated around the membrane, generating an
adsorption layer, which is significant because of the near-criticality (Fig.~\ref{fig:sample}).
The osmotic pressure caused by the concentration gradient should influence the membrane motion.  
In Ref.~\onlinecite{pre}, the present author studied the influence by assuming
the Gaussian free-energy functional and weak preferential attraction, and 
showed that
a new term is added to the restoring force of the membrane and 
tends to suppress the mean square amplitude of the membrane. We abbreviate
this equal-time correlation as MSA. 
We here calculate the MSA
of a tensionless membrane under more general conditions,
considering that the preferential attraction is not always 
weak in practice and that 
the Gaussian model ceases to be valid as the mixture becomes
sufficiently close to the critical point. 
\\

The following assumptions are shared by Ref.~\onlinecite{pre} and this study.
The temperature is homogeneous and constant.
The membrane, made up of a single component without electric charge, 
is immersed in an incompressible
binary fluid mixture without ions, and is regarded 
as a thin film fluctuating around a plane.  The mixture has the same properties on
both sides of the membrane.  Far from the membrane, it is  
in the homogeneous phase near the demixing critical point.
The preferential attraction comes from a short-range interaction and is represented by
the surface field.  As mentioned in Refs.~\onlinecite{pre} and \onlinecite{relax},  
these assumptions would be satisfied experimentally by means of
a membrane with outward nonpolar chains in some organic binary mixture, for example.
Focusing on modes with wavelength longer than the correlation length,
we calculate fields of the mixture within the linear approximation
with respect to the amplitude. 
In Ref.~\onlinecite{pre},  
the governing equations in this approximation were derived from the Gaussian free-energy functional
and solved approximately on the assumption of a sufficiently weak surface field.
The conditions under which the result in this previous study is valid were left quantitatively undetermined
in terms of the wavenumber of the undulation mode and the temperature.   
In the present study, we make them explicit and elucidate what happens when
they are not satisfied.  \\

An equal-time correlation is obviously
independent of the dissipative part of the dynamics.  However,
the reversible part can affect an equal-time correlation.   Let us illustrate this effect with a simple example, where
a colloidal sphere (mass $m_{\rm p}$) is trapped by an external potential in a one-component fluid.
Describing the small oscillation about the equilibrium point, we apply
the equipartition theorem to calculate the mean square velocity \cite{kubo,relax,sphere}.
This equal-time correlation 
is given by $k_{\rm B}T/(m_{\rm p}+m_{\rm ind})$, where $k_{\rm B}$ and $T$ respectively denote the Boltzmann constant
and temperature.  The induced mass $m_{\rm ind}$ is 
half the mass of the displaced fluid \cite{lamb,grim}.
To show this, we have only to use the reversible part of the
hydrodynamics, although the real dynamics is dissipative. 
In our problem, 
besides driving flow, the membrane motion changes the local concentration gradient in the mixture
and deforms the adsorption layer dynamically.
The equipartition theorem we use, which is shown later by Eq.~(\ref{eqn:equip}),
involves the induced-mass density and normal-mode frequency.
We formulate the reversible part of the hydrodynamics,
as in Refs.~\onlinecite{pre} and \onlinecite{sphere}, by using 
a coarse-grained free-energy functional \cite{ofk,furu,wetdrop,visc,wetraft}. 
Considering imaginary periodic motion about the equilibrium point, 
we can calculate the MSA of a real system.
\\

Our formulation is shown in Sect.~\ref{sec:form}, where the contents up to Eq.~(\ref{eqn:newouter3})
are essentially described in Ref.~\onlinecite{pre}.  
We show and discuss our results within the Gaussian model in Sect.~\ref{sec:gauss},
where no assumption is imposed on the magnitude of the surface field.
Some details of the calculation procedure are relegated to appendices.
The result in Ref.~\onlinecite{pre} turns out to be valid only
for undulation modes with large wavenumber.   For modes with small wavenumber, 
the numerical factor of the additional restoring force approaches half that in the previous result
because of decreased hydrodynamic effects.  
In the intermediate range of wavenumber, the induced-mass density increases to suppress
the MSA considerably.  A model more general than the Gaussian model is
applied in Sect.~\ref{sec:general}, where
we assume a critical composition far from the membrane to consider only the modes with small wavenumber.
Beyond the regime of the Gaussian model, it is found  that
the additional restoring force does not increase with the
correlation length far from the membrane.  This is not expected from
the results in the Gaussian model, where the correlation length is
homogeneous.  The result beyond its regime 
is roughly given by the corresponding result in the Gaussian model
if the correlation length is interpreted as one close to the membrane,
which reaches a plateau as the temperature approaches the critical value.
Our numerical results are obtained with the aid of Mathematica (Wolfram Research).  
Section \ref{sec:dis} gives a summary and outlook.  

\section{\label{sec:form}Formulation}
\subsection{Equations for the mixture \label{sec:form1}}
In the mixture, the mass-density difference between the two components, denoted by 
$\varphi$, depends on the position $\bm{r}$.  
The preferential attraction is represented by
the potential $f_{{\rm s}}$ determined by the value of $\varphi$
very near the membrane.
Near the critical point, the $\varphi$-dependent part of the free-energy functional is  
given by \cite{Cahn}
\begin{eqnarray}&&
\int_{C^{\rm e}} d\bm{r}\ 
\left[f(\varphi(\bm{r}))+{1\over 2}M(\varphi(\bm{r}))
\left\vert\nabla\varphi(\bm{r})\right\vert^2\right]\nonumber\\
&&\qquad\quad+\int_{\partial C }dS\ 
f_{{\rm s}}(\varphi(\bm{r}))
\ ,\label{eqn:glw}\end{eqnarray}
which has been coarse-grained up to the local correlation length. 
The first integral is the volume integral over the mixture regions
on both sides of the membrane ($C^{\rm e}$); $f$ and $M(>0)$ are functions of $\varphi$.
The second integral is the surface integral
over the interfaces on both sides of the membrane ($\partial C$).  
We assume $f_{\rm s}$
to be a linear function; $-df_{\rm s}(\varphi)/(d\varphi)\equiv h$ is called the surface field. 
If the mixture is not too close to the critical point,
$f$ and $M$ can be regarded as
a quadratic function and a positive constant, respectively.  Then, we have the
Gaussian model.  \\

\begin{figure}[b]
\includegraphics[width=8cm]{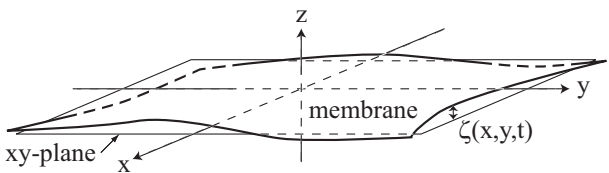}
\caption{\label{fig:memb} Schematically drawn fluid membrane, made up of a single component, fluctuating around the $xy$-plane.
The ambient mixture has the same properties  
on both sides of the membrane. }
\end{figure}

The hydrodynamics derived from the bulk part of Eq.~(\ref{eqn:glw})
is well known in the model H \cite{Hohenberg, Onukibook},  which was used in
studying the relaxation of the two-time correlation in a near-critical fluid.
We here utilize only its reversible part
for calculating the equal-time correlation.  
The Cartesian coordinate system $(x,y,z)$ is set so that
the membrane fluctuates around the $xy$-plane (Fig.~\ref{fig:memb}).
The $z$ coordinate of
the membrane is referred to as $\zeta$, which is a function of $x, y,$ and the time $t$. 
The time dependence of each field is considered below. 
The chemical potential conjugate to $\varphi$ is given by
\begin{equation}
\mu(\bm{r},t)=f'(\varphi(\bm{r},t))-{1\over 2}M'|\nabla\varphi|^2
-M\Delta\varphi(\bm{r},t) \ .
\label{eqn:hatmudef}
\end{equation}
Hereafter, a prime indicates the derivative with respect to the
variable, {\it e.g.\/}, $f'=df/(d\varphi)$ and $M'=dM/(d\varphi)$, while a double prime
means the second derivative.
We write $\bm{n}$ for the unit vector which is
normal to the membrane and directed towards the
positive-$z$ side.  The local equilibrium at the interface gives the boundary condition \cite{ofk,jans,panof}
\begin{equation}
\pm M\bm{ n}\cdot \nabla 
\varphi =-h \quad {\rm as}\ z\to \zeta \pm
\label{eqn:phisurface}\ ,\end{equation}
where $z\to \zeta +(-)$ means that $z$ approaches $\zeta$ with
$z>\zeta\ (<\zeta)$ maintained. 
We write $\bm{V}$ for the velocity field in the mixture.  
Its mass density, denoted by $\rho$, is regarded as constant \cite{broc,seif}.  
The incompressibility is represented by
$\nabla\cdot \bm{ V}=0$, 
while the equation of motion is given by
\begin{equation}
\rho{\partial \bm{ V}\over \partial t}
=-\nabla {p}-\varphi \nabla\mu \ ,
\label{eqn:3Ddyn}\end{equation}
where the convective term is neglected in anticipation of the
later linear approximation.    
We need not assume viscosity to calculate the equal-time correlation,
as mentioned in Sect.~\ref{sec:intro}.
The scalar pressure ${p}$ 
exists here to keep the incompressibility.
The term $-\varphi\nabla\mu$ originates from
the reversible part of the pressure tensor, which is given 
by Eq.~(5) of Ref.~\onlinecite{pre} and contains the osmotic pressure. 
This part can be derived from
the first term of Eq.~(\ref{eqn:glw}) \cite{Onukibook, ofk}.
At the membrane, the velocity field follows the boundary condition for an inviscid fluid \cite{bound}.
This condition requires only the normal component 
to be continuous within the linear approximation, as shown by Eq.~(\ref{eqn:vzl}) below. 
\\

Far from the membrane, 
$\bm{ V}$ vanishes, and $\varphi$, $\mu$, and ${p}$ are constant.
Their constant values are respectively denoted by $\varphi_\infty$, $\mu^{(0)}\equiv f'(\varphi_\infty)$, 
and ${p}^{(0)}$, which are shared by the mixture regions on both sides of the membrane.
Because the diffusive flux between the two components is proportional to
the gradient of $\mu$,
the mass conservation of each component leads to  
\begin{equation}
{\partial\varphi\over\partial t}
=-\bm{ V}\cdot \nabla\varphi+ \nabla\cdot L \nabla \mu\ ,
\label{eqn:phidyn}\end{equation}
where the Onsager coefficient $L$ is a positive function of $\varphi$. 
Assuming that the diffusion flux cannot pass across the membrane
leads to
\begin{equation}
\bm{ n}\cdot L\nabla\mu=0\quad {\rm as}\ z\to \zeta \pm
\ .\label{eqn:musurface}\end{equation}
The diffusion should not be involved in the equal-time correlation, and
we will take the limit of $L\to 0+$ later.  This limit gives rise to an unfamiliar boundary layer 
of the chemical potential
unless $h$ vanishes; $L$ is associated with the highest-order derivative in Eq.~(\ref{eqn:phidyn}) \cite{bender}.
 (In passing, the dissipative dynamics based on the Gaussian model is considered 
together with Eqs.~(\ref{eqn:hatmudef}), (\ref{eqn:phisurface}), (\ref{eqn:phidyn}), and (\ref{eqn:musurface}) in Refs.~\onlinecite{relax},
\onlinecite{ofk}, and \onlinecite{wetdrop}--\onlinecite{wetraft}.)
We later consider equations of the mixture fields on the positive-$z$ side; the solution 
outside the boundary layer 
is referred to as the outer solution. 

\subsection{Expansion with respect to the undulation amplitude}
We here prepare to perform the calculation within the linear approximation with respect to the
amplitude.  Introducing a dimensionless parameter
$\varepsilon$, we define nonzero $\zeta^{(1)}$ so that 
$\zeta(\bm{x},t) =\varepsilon \zeta ^{(1)}(\bm{x},t)$, 
where $\bm{x}$ represents a position on the membrane
and has coordinates $(x,y)$.  We define the mean curvature of the membrane $H$ 
so that it is positive when the center of the curvature lies on the side
towards which $\bm{ n}$ is directed. 
Assuming that the spontaneous curvature vanishes, we 
write $c_{\rm b}H^2$ for the bending energy per unit area of the membrane, where 
 $c_{\rm b}$ is the bending rigidity \cite{canh,helfbend}.
Let us write $\bm{ v}(\bm{x}, t)$ for the velocity field of the membrane and
$\bm{F}(\bm{x}, t)$ for the stress exerted on the membrane by the mixture.
Assuming the membrane to be compressible, we write
$\rho_{\rm m}(\bm{x} ,t)$ for the membrane mass per unit area and
$p_{\rm m}(\bm{x},t)$ for its in-plane pressure field.
Neglecting the membrane viscosity,
we can describe the mass conservation and momentum conservation of the membrane
\cite{seif,physica,myjcp,powe}.
Up to the order of $\varepsilon$, the equation of the normal motion is separated from
that of the tangential motion, as discussed around Eq.~(70) of
Ref.~\onlinecite{pre}, and is given by
\begin{equation}
\rho_{\rm m}{\partial v_z \over \partial t}
=F_z+F_r-2Hp_{\rm m}\label{eqn:memz}
\ ,\end{equation}
where $F_r\equiv -c_{\rm b}\left[\partial^2/(\partial x^2)+\partial^2/(\partial y^2)\right]H$ 
represents the restoring force against bending \cite{ouyang,ouyang2}.
The right-hand side (rhs) of Eq.~(\ref{eqn:memz})
gives the total restoring force.    \\

The equilibrium state
for the planar membrane is the reference state, {\it or\/} unperturbed state, with $\varepsilon=0$, where  
$\mu$ is homogeneous and so is ${p}$ 
because of Eq.~(\ref{eqn:3Ddyn}) \cite{ofk}. They are respectively given by the constants
$\mu^{(0)}$ and ${p}^{(0)}$.  For $\varphi$ in the unperturbed state, we write
$\varphi^{(0)}$, which depends only on $z$ and is even with respect to $z$.  Up to the order of $\varepsilon$, 
we expand the fields as
\begin{eqnarray}
& &\varphi(\bm{ r},t)=\varphi^{(0)}(z)+\varepsilon 
\varphi^{(1)}(\bm{ r},t)\ , \nonumber\\ &&
\mu(\bm{ r},t)=\mu^{(0)}+\varepsilon \mu^{(1)}(\bm{ r},t)\ ,
\nonumber\\
& &{p}(\bm{ r},t)={p}^{(0)}+\varepsilon {p}^{(1)}(\bm{ r},t) 
\ ,\nonumber\\&& {\rm and }\ \bm{ V}(\bm{ r},t)=\varepsilon \bm{ V}^{(1)}(\bm{ r},t)
\ ,\label{eqn:perexp}\end{eqnarray}
whereby the fields with the
superscript $^{(1)}$ are defined.  These fields 
vanish far from the membrane.  
The boundary condition for an inviscid fluid gives
\begin{equation}
\lim_{z\to 0\pm }{V}_z^{(1)}={v}_z^{(1)}= {\partial {\zeta}^{(1)} \over\partial t}\ ,
\label{eqn:vzl}\end{equation}
while Eq.~(\ref{eqn:musurface}) gives
\begin{equation}
L{\partial \mu^{(1)}\over \partial z}\to 0\quad {\rm as}\ z\to 0\pm 
\ .\label{eqn:musurface2}\end{equation}
The boundary condition, Eq.~(\ref{eqn:phisurface}), should hold 
in both the unperturbed and perturbed states.  If $M$ is a constant \cite{even}, we thus have
\begin{equation}
{\partial \varphi^{(1)}\over\partial z}
+{\varphi^{(0)}}''\zeta^{(1)}=0\quad {\rm as}\ z\to 0\pm
\ ,\label{eqn:bcphi}\end{equation}
which is the same as Eq.~(29) of Ref.~\onlinecite{pre}.  
\\

We write $(\bm{x}, z)$ for $\bm{r}=(x,y,z)$.   In the directions of
$x$ and $y$,  we impose the periodic boundary condition and
add an overhat  to the Fourier transform, {\it e.g.\/},  
\begin{equation}
{\hat p}^{(1)}(\bm{ k},z,t)\equiv {1\over  l_{\rm p}^2}\int_{0}^{l_{\rm p}}dx
\int_{0}^{l_{\rm p}}dy\   
p^{(1)}(\bm{x},z,t)e^{-i\bm{k}\cdot\bm{x}}
\ ,\label{eqn:four}\end{equation} 
where $l_{\rm p}$ is the period and $\bm{ k}$ represents $(k_x,k_y)$ with
$l_{\rm p}k_x/(2\pi)$ and $l_{\rm p}k_y/(2\pi)$ being
integers.  We assume that $k\equiv \left|\bm{k}\right|$ does not vanish because 
the translational shift is of no interest; $l_{\rm p}$ can be regarded as representing the membrane size.
We add an overtilde to the further Fourier transform with respect to $t$, {\it  e.g.\/},
\begin{equation}
{\tilde p}^{(1)}(\bm{ k},z,\omega)
={1\over 2\pi} \int_{-\infty}^\infty dt\  
{\hat p}^{(1)}(\bm{ k},z,t)
e^{i\omega t}\ .\label{eqn:fourtime}
\end{equation}
As in Eq.~(\ref{eqn:perexp}), 
we expand the membranous fields; $\rho_{\rm m}^{(0)}$
denotes the term of $\rho_{\rm m}$ independent of $\varepsilon$, and
$\varepsilon F_z^{(1)}$ equals $F_z$ up to the order of $\varepsilon$.   
Equations (\ref{eqn:memz}) and (\ref{eqn:vzl}) yield
\begin{equation}
-\rho_{\rm m}^{(0)}\omega^2 {\tilde \zeta}^{(1)}={\tilde F}^{(1)}_z
-\left({c_{\rm b}k^4\over 2}-p^{(0)}_{\rm m} k^2\right) {\tilde \zeta}^{(1)}
\label{eqn:maku}\end{equation}
because ${\tilde H}$ equals $-\varepsilon k^2{\tilde \zeta}^{(1)}/2$
up to the order of $\varepsilon$. 
We write $\varphi^{(0)}(0+)$ for $\varphi^{(0)}(z)$ in the limit of $z\to 0+$.
As shown by Eq.~(71) of Ref.~\onlinecite{pre},
the mean lateral tension is given by
 \begin{equation}
\sigma_{\rm l} \equiv -p^{(0)}_{\rm m}+2f_{\rm s}(\varphi^{(0)}(0+))
\ ,\label{eqn:lat}\end{equation} 
which should vanish 
for a membrane suspended freely in the mixture. 
\\

From Eq.~(\ref{eqn:3Ddyn}) with the incompressibility condition, 
we can derive
\begin{equation}
\left({\partial^2\over\partial z^2}-k^2\right) 
{\tilde V}_z^{(1)}=-{ik^2\over \rho\omega}{{\varphi}^{(0)}}'
{\tilde\mu}^{(1)}\ .\label{eqn:eqforvz}
\end{equation}
This is the same as  Eq.~(41) of Ref.~\onlinecite{pre}, where the
derivation is described in more detail; 
the component of $\tilde{\bm{V}}^{(1)}$ perpendicular to $\bm{k}$ and to the $z$ direction is not associated with the
membrane motion.  
Below, the subscript $_{\rm out}$ indicates the outer solution mentioned at the end of Sect.~\ref{sec:form1};
the outer solution of $\varphi^{(1)}$ is denoted by $\varphi^{(1)}_{\rm out}$, for example.
Outside the boundary layer, we can neglect the second term on the rhs of Eq.~(\ref{eqn:phidyn})
in the limit of $L\to 0+$, as discussed below Eq.~(59) of Ref.~\onlinecite{pre} and Eq.~(45) of Ref.~\onlinecite{sphere}. 
The Fourier transform gives
\begin{equation}
i\omega{\tilde\varphi}^{(1)}_{\rm out}={\tilde V}^{(1)}_{{\rm out}\ z} {\varphi^{(0)}}'(z)
\ .\label{eqn:newouter}\end{equation}

\subsection{Gaussian model \label{sec:formg}}
We here describe the Gaussian model,
mentioned below Eq.~(\ref{eqn:glw}), in more detail.  In this model,
$M$ is a positive constant and $f(\varphi)$ is assumed to be 
given by \begin{equation}
{m \over 2} \left(\varphi-\varphi_\infty\right)^2
+\mu^{(0)}\left(\varphi-\varphi_\infty\right)
\ ,\label{eqn:gauss}\end{equation}
with $m$ being defined as a positive constant.  
Picking up the terms independent of $\varepsilon$ from Eqs.~(\ref{eqn:hatmudef})
and (\ref{eqn:phisurface}), we obtain
\begin{equation}
\varphi^{(0)}(z)
=\varphi_{\infty} + {h\xi_{\rm c}\over M} e^{- |z|/\xi_{\rm c}}
\label{eqn:phizero}\ ,\end{equation}
as shown by Eq.~(27) of  Ref.~\onlinecite{pre}.  Here,
$\xi_{{\rm c}} \equiv \sqrt{{M/m}}$ 
is the correlation length far from the membrane.  It is typically up to several nanometers
in the regime of the Gaussian model, as discussed in Sect.~\ref{sec:genfree}.
The profile above is the same as that obtained in a mixture in contact with a flat wall; it
is essentially calculated in Ref.~\onlinecite{Cahn} and is also 
mentioned in Sect.~IIA of Ref.~\onlinecite{binder}.  \\

Extracting the terms at the order of $\varepsilon$ from Eq.~(\ref{eqn:hatmudef}), we have
\begin{equation}
\left[M\left({\partial^2\over\partial z^2}-k^2\right)-m\right]{\tilde \varphi}^{(1)}
=-{\tilde \mu}^{(1)}
\ .\label{eqn:newouter3}\end{equation}
Equations (\ref{eqn:eqforvz}) and (\ref{eqn:newouter3}) remain valid if ${\tilde V}^{(1)}_{z}$, ${\tilde \mu}^{(1)}$, and
${\tilde \varphi}^{(1)}$ are replaced by their respective outer solutions. 
Substituting Eq.~(\ref{eqn:newouter}) into 
Eq.~(\ref{eqn:newouter3}), we use Eqs.~(\ref{eqn:eqforvz}) and (\ref{eqn:phizero})
to derive a closed equation for ${\tilde V}^{(1)}_{{\rm out}\ z}$ and find that its solution is
given in terms of Gauss' hypergeometric series, as shown in  
Sect.~\ref{sec:gaussianproc} and
Appendix \ref{app:a}.  Using this solution, which is linearly related with ${\tilde\zeta}^{(1)}$
through the boundary condition at the membrane,
we rewrite ${\tilde F}^{(1)}_z$ of Eq.~(\ref{eqn:maku}) in terms of ${\tilde\zeta}^{(1)}$
to calculate the normal-mode frequency in Sect.~\ref{sec:gaussiannormal}.
In this calculation, for which some equations are prepared below, 
we have only to consider the fields on the positive-$z$ side
because of the symmetry. 
Suppose $h\ne 0$.
We write $\triangle(>0)$ for the thickness of the boundary layer, which 
decreases to zero as $L$ decreases to zero \cite{comBL}. 
Considering Eqs.~(\ref{eqn:bcphi}) and (\ref{eqn:newouter3}), ${\tilde \mu}^{(1)}$ and
$\partial^2{\tilde \varphi}^{(1)}/(\partial z^2)$
are likewise singular in the boundary layer in the limit of $L\to 0+$, while $\partial {\tilde \varphi}^{(1)}/(\partial z)$ remains finite there.
Integrating Eq.~(\ref{eqn:newouter3}) from $z=0$ to $\triangle$, we find 
\begin{equation}
\left[ M {\partial \over \partial z} {\tilde \varphi}^{(1)}(\bm{k},z,\omega)
\right]_0^\triangle= 
-\int_0^\triangle dz\ {\tilde \mu}^{(1)}(\bm{k},z,\omega)
\label{eqn:mu1limitkantan}\end{equation}
in the limit of $L\to 0+$, {\it i.e.\/}, $\triangle\to 0+$, where
the left-hand side (lhs) means the value of the term in the braces at $z=0$ subtracted from
the value at $z=\triangle$. Similarly,
the rhs of Eq.~(\ref{eqn:eqforvz}) is affected by the singular behavior of ${\tilde \mu}^{(1)}$
in the limit of $L\to 0+$, but $\partial {\tilde V}^{(1)}_z/(\partial z)$ remains finite in the boundary layer.
Taking the limit of $L\to 0+$ for the integral of this derivative across the boundary layer, we obtain
\begin{equation}
\lim_{z\to 0+} {\tilde V}^{(1)}_{z}
 =\lim_{z\to 0+} {\tilde V}^{(1)}_{{\rm out}\ z}\ ,
\label{eqn:Voutlim}\end{equation}
which is combined with Eq.~(\ref{eqn:vzl}) to give a boundary condition for ${\tilde V}^{(1)}_{{\rm out}\ z}$.
Applying this procedure to $\partial {\tilde \varphi}^{(1)}/(\partial z)$
and using Eqs.~(\ref{eqn:phisurface}), (\ref{eqn:vzl}), 
(\ref{eqn:newouter}), and (\ref{eqn:Voutlim}),  we obtain 
\begin{equation}
\lim_{z\to 0+} {\tilde \varphi}^{(1)}=\lim_{z\to 0+} {\tilde \varphi}^{(1)}_{\rm out}
= {h\over M}{\tilde \zeta}^{(1)} \label{eqn:phiz0}\end{equation}
in the limit of $L\to 0+$.
Using Eqs.~(\ref{eqn:hatmudef})--(\ref{eqn:3Ddyn}), (\ref{eqn:bcphi}), 
and (\ref{eqn:phiz0}),  
we have
\begin{equation}
{\tilde F}_z^{(1)}= -2k^2{\tilde \zeta}^{(1)}f_{\rm s}(\varphi^{(0)}(0+))
-2\rho {i\omega\over k^2} \lim_{z\to 0+} {\partial {\tilde V}^{(1)}_z\over\partial z} 
\label{eqn:tilFz}\ ,\end{equation} 
because ${\tilde V}_z^{(1)}$ is even with respect to $z$ \cite{com1}.  
We can rewrite the last term above in terms of $\partial {\tilde V}^{(1)}_{{\rm out}\ z}/(\partial z)$
at $z\to 0+$ by using Eq.~(\ref{eqn:mu1limitkantan}), as shown by Eq.~(\ref{eqn:partUU}).

\section{Results in the Gaussian Model\label{sec:gauss}}
An outline of our procedure in the Gaussian model is described in this paragraph;
details are shown in Sect.~\ref{sec:gaussianproc}.
In each normal mode of the small oscillation about the
equilibrium point, we can write the total kinetic energy as
\begin{equation}
{\varepsilon^2 l_{\rm p}^2\over 2} \left[
\rho_{\rm m}^{(0)}
\left| {\partial {\hat \zeta}^{(1)}(\bm{k}, t)\over \partial t}
\right|^2+  \rho\int_{-\infty}^\infty dz\ \left|
 \hat{\bm{V}}^{(1)}(\bm{k}, z, t)\right|^2 \right]
\ .\label{eqn:kineticenergy}\end{equation}
Considering the statement below Eq.~(\ref{eqn:eqforvz}), 
we can rewrite the integrand above
in terms of ${\hat V}^{(1)}_z$ and $\bm{k}\cdot \hat{\bm{V}}^{(1)}$.
They are related with each other through the incompressibility condition, as shown by Eq.~(40) of Ref.~\onlinecite{pre},
and ${\hat V}^{(1)}_z$ can be replaced
by ${\hat V}^{(1)}_{{\rm out} \ z}$ in the integrand
because of Eq.~(\ref{eqn:Voutlim}) and the statement above it.  Furthermore, ${\hat V}^{(1)}_{{\rm out} \ z}$ is
related linearly with $\partial {\hat \zeta}^{(1)}/\left(\partial t\right)$.
Thus, in each normal mode, we can define $\rho^{({\rm eff})}_k$ so that
Eq.~(\ref{eqn:kineticenergy}) is rewritten as
\begin{equation}
{\varepsilon^2 l_{\rm p}^2\over 2} 
\rho^{({\rm eff})}_k
\left| {\partial {\hat \zeta}^{(1)}(\bm{k}, t)\over \partial t}
\right|^2
\ .\label{eqn:kinetic}\end{equation}
The induced-mass density is given by $\rho^{({\rm eff})}_k-\rho_{\rm m}^{(0)}$.  
In each normal mode, the total potential energy equals 
\begin{equation}
{\varepsilon^2 l_{\rm p}^2\over 2} 
\rho^{({\rm eff})}_k \omega_*^2
\left| {\hat \zeta}^{(1)}(\bm{k}, t)
\right|^2
\ ,\label{eqn:potential}\end{equation}
where $\omega_*$ is the normal-mode frequency.
When a single normal-mode frequency occurs for a given wavenumber vector, 
the MSA is given by
\begin{equation}
\langle {\hat \zeta}(\bm{k},t) {\hat \zeta}(\bm{k}',t)\rangle ={\delta_{\bm{k}, -\bm{k}' }k_{\rm B}T\over l_{\rm p}^2}\times
{1\over \rho^{({\rm eff})}_k \omega_*^2 }
\ ,\label{eqn:equip}\end{equation}
where $\langle\cdots\rangle$ indicates the equilibrium average. 
More than one normal-mode frequency can occur, as shown in
Sect.~\ref{sec:gaussiannormal}.   Results for its different numbers
are separately shown in Sects.~\ref{sec:gaussiansingle} and \ref{sec:gaussianthree}. 
Details of the calculation procedure are described in Sect.~\ref{sec:gaussianproc}.

\subsection{Normal-mode frequency\label{sec:gaussiannormal}}
The dimensionless surface field and frequency are respectively defined as
\begin{equation}
\lambda\equiv {h\xi_{\rm c}^{3/2} \over \sqrt{c_{\rm b}M}}\quad  {\rm and}\quad 
\Omega \equiv  {\omega \sqrt{\rho \xi_{\rm c}^3 }\over 
\lambda k\sqrt{c_{\rm b}}}=
 {\omega \sqrt{\rho M} \over hk}
\ .\label{eqn:Lamgauss}\end{equation}
We use $K\equiv k\xi_{\rm c}$, which should be much smaller than unity in
our hydrodynamic formulation based on the coarse-grained free-energy functional,
Eq.~(\ref{eqn:glw}).  
For a tensionless membrane, we can rewrite Eq.~(\ref{eqn:maku}) as
\begin{equation}
\left(\rho_{\rm m}^{(0)}+ {2\rho \over k} \right) \omega^2 {\tilde \zeta}^{(1)}
={c_{\rm b} \over \xi_{\rm c}^4}  \left[ {K^4 \over 2}
+2 \lambda^2 {\left(1+d_1\right)  K^2\over K+1}\right]  {\tilde \zeta}^{(1)}\ ,
\label{eqn:perpendrep}\end{equation} 
whereby $d_1$ is defined; $d_1$ vanishes in the previous result of Ref.~\onlinecite{pre}.
As shown in Appendix \ref{app:stress},
we find that $d_1$ can be given in terms of Gauss' hypergeometric series.
We write $d_1(K, \Omega^2)$ to clarify its dependence on $K$ and $\Omega^2$.
Typically, we have $\rho\approx 1\ $g/cm$^3$ and $\rho^{(0)}_{\rm m}\approx 10^{-7}\ $g/cm$^2$. 
For the values of $k$ of our interest, we can neglect $\rho^{(0)}_{\rm m}$ on the lhs of Eq.~(\ref{eqn:perpendrep}).\\

\begin{figure}
\includegraphics[width=8cm]{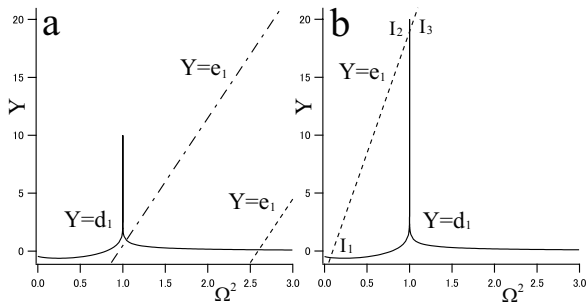}
\caption{\label{fig:cald}  Graphs of $d_1$ and $e_1$.
(a) The solid curve represents $Y=d_1(0.1, \Omega^2)$ calculated from Eq.~(\ref{eqn:deeone}).  
The dashed line and dashed-dotted line represent $Y=e_1(0.1, \Omega^2)$ for $\lambda=0.01$
and $0.017$, respectively.  The slope 
of Eq.~(\ref{eqn:e1func}) is approximately $1/K$.  
(b) The solid curve and the dashed line
respectively represent $Y=d_1(K, \Omega^2)$ and $Y=e_1(K, \Omega^2)$ for $(\lambda, K)=(0.025, 0.05)$; 
the intersections are named as $I_1$, $I_2$, and $I_3$.
When $K\ll 1$ is given, we have $d_1(K, \Omega^2)\approx -1/2$ in the limit of $\Omega^2\to 0+$, 
as shown at the end of Appendix \ref{app:stress}.  }
\end{figure} 

We suppose $h\ne 0$.
For a given value of $K(\ne 0)$, the normal-mode frequency is determined
by the solution of Eq.~(\ref{eqn:perpendrep}).  Manipulating this equation, we
find the solution to be given by the 
intersection of the curve $Y=d_1(K,\Omega^2)$ and
\begin{equation}
Y={K+1\over K} \Omega^2-{K^2\left(K+1\right) \over 4\lambda^2 }-1
\ .\label{eqn:e1func}\end{equation}
The rhs above is denoted by $e_1(K,\Omega^2)$, which also depends on $\lambda^2$.  
The $Y$ and $\Omega^2$ intercepts
of the line $Y=e_1$ are respectively given by
\begin{equation}
-{K^2(K+1)\over 4\lambda^2 }-1\quad {\rm and}\quad {K^3\over 4\lambda^2 }+{K\over K+1}
\ ,\label{eqn:intercept}\end{equation}
which respectively tend to $-1$ and $0$ as $K$ decreases.  For $\Omega$ at the intersection, 
we write $\Omega_*$, which is
$\Omega$ of  Eq.~(\ref{eqn:Lamgauss}) at $\omega=\omega_*$.  \\

The dashed line in Fig.~\ref{fig:cald}(a), representing $Y=e_1$ with $K^3=10 \lambda^2$, 
has one intersection with $Y=d_1$ 
in the region where $d_1$ almost vanishes.
In this case, we can neglect $d_1$ in Eq.~(\ref{eqn:perpendrep})
to obtain the normal mode frequency.  This leads to
Eq.~(73) of Ref.~\onlinecite{pre} with $\sigma_{\rm l}$ put equal to zero.
Judging from the second entry of Eq.~(\ref{eqn:intercept}), $K^3/(4\lambda^2)$
roughly gives the $\Omega^2$ intercept of the line $Y=e_1$. 
When a single intersection occurs, 
$\Omega_*^2$ is larger than unity and is roughly equal to
$K^3/(4\lambda^2)$.  
As $\lambda$ increases, the line $Y=e_1$ shifts translationally towards the left-hand side.
The dashed-dotted line in Fig.~\ref{fig:cald}(a) has
$K^3/(4\lambda^2)=8.7\times 10^{-1}$; it still has a single intersection, where 
$d_1$ is no longer negligible.  
As $K^3/\lambda^2$ decreases further, three intersections occur because 
$e_1(K, 1)<1/K$.
An example of this is shown in Fig.~\ref{fig:cald}(b), where $K^3\ll 4\lambda^2 $ holds.
The three intersections are referred to as  
$I_1$, $I_2$, and $I_3$ in order of proximity to the $Y$ axis, 
although $I_2$ and $I_3$ cannot be distinguished in the figure.  Thus,
when $K^3/\lambda^2$ is sufficiently small, 
three normal modes occur per wavenumber vector $\bm{k}$.

\begin{figure}
\includegraphics[width=8cm]{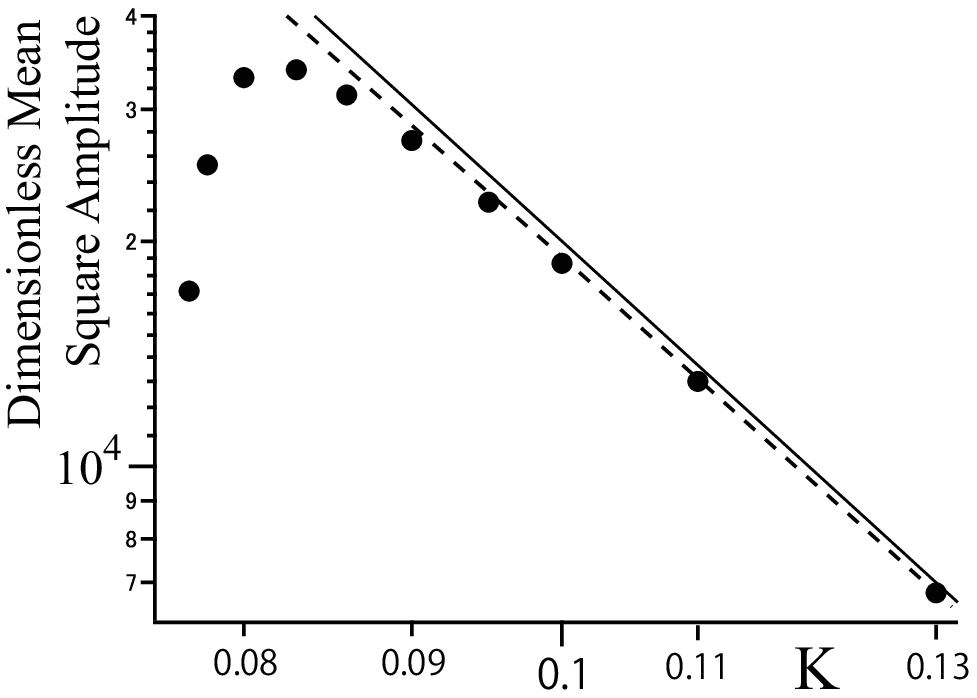}
\caption{\label{fig:zoku4}  Graphical representation of the MSA 
in the range of $K$ causing a single intersection.
We use Eqs.~(\ref{eqn:equip}) and (\ref{eqn:kikamoto}) to calculate
$\langle {\hat \zeta}(\bm{ k},t){\hat \zeta}(-\bm{ k},t)
\rangle$ numerically.  Dividing it 
by $k_{\rm B}T_{\rm c} \xi_{\rm c}^4 /\left(l_{\rm p}^2c_{\rm b}\right)$, we nondimensionalize 
the MSA and plot the quotient (circles). 
The dashed line represents the reciprocal of the sum in the braces of Eq.~(\ref{eqn:main}).  
We use $\lambda=1.16\times 10^{-2}$, which arises at
$\xi_{\rm c}=3$ nm for the material constants mentioned in Sect.~\ref{sec:genfree}; 
a single intersection occurs for $K>7.6\times 10^{-2}$.   
  The solid line represents $2/K^4$, which gives the
dimensionless MSA for $h=0$.  }
\end{figure}

\subsection{Case of a single intersection\label{sec:gaussiansingle}}
Using Eq.~(\ref{eqn:equip}), we numerically calculate the MSA when a single intersection occurs
because of large $K^3/(4\lambda^2)$.  
 The results are plotted with
circles in Fig.~\ref{fig:zoku4}. 
The ratio $K^3/\lambda^2$ equals $c_{\rm b}Mk^3/h^2$, which is independent of
$\xi_{\rm c}$.
When the ratio is large enough to make $d_1$ negligible at the single intersection,
we can estimate the MSA as follows.  Then,
the induced-mass density is found to be approximately equal to $2\rho/k$, as
mentioned in Sect.~\ref{sec:gaussianproc}. 
Thus, for a tensionless membrane, we use $d_1\approx 0$ in Eqs.~(\ref{eqn:equip}) and (\ref{eqn:perpendrep}) to obtain
\begin{eqnarray}
&&\langle {\hat \zeta}(\bm{ k},t){\hat \zeta}(\bm{ k}',t)\rangle \nonumber\\ &&\qquad\qquad
\approx \delta_{\bm{ k},-\bm{ k}'}{k_{\rm B}T_{\rm c} \xi_{\rm c}^4  \over l_{\rm p}^2{c_{\rm b}}}
 \left[ {K^4\over 2}
+2 \lambda^2 {K^2\over K+1}   \right]^{-1}\label{eqn:main}\\
&&\qquad\qquad=\delta_{\bm{ k},-\bm{ k}'}{k_{\rm B}T_{\rm c} \over l_{\rm p}^2}
\left[ {c_{\rm b}k^4\over 2} + {2 h^2 \over M\xi_{\rm c}} {K^2\over K+1} \right]^{-1}\ .
\label{eqn:mainato}\end{eqnarray} 
With $\sigma_{\rm l} k^2$ being added in the braces, Eq.~(\ref{eqn:mainato}) is 
the same as Eq.~(75) of Ref.~\onlinecite{pre}, which is obtained for sufficiently weak preferential attraction.
The first and second terms in the braces of Eq.~(\ref{eqn:mainato}), after multiplied by $-{\hat\zeta}$,
give the restoring forces due to the bending rigidity and
ambient near-criticality, respectively.
For $K\stackrel{>}{\sim} 0.09$, {\it i.e.\/},  $K^3/(4\lambda^2)\stackrel{>}{\sim}1.35$, in Fig.~\ref{fig:zoku4},
Eq.~(\ref{eqn:main}) gives a good approximation to the circles.
 They are slightly below the solid line; 
the ambient near-criticality suppresses
the MSA slightly.   This slight suppression is because the large $K^3/\lambda^2$
makes the first term predominant in the braces of Eq.~(\ref{eqn:mainato}). \\

The second term on the rhs of Eq.~(\ref{eqn:3Ddyn})
depends on the concentration gradient because of Eq.~(\ref{eqn:hatmudef}).
In the small oscillation about the equilibrium point, 
this term generates a restoring force of the periodic motion of the mixture,
judging from the suppression effect.
It appears as if harmonic oscillators were distributed over the mixture.
Their natural frequencies vary according to the distance from the membrane because of the 
varying concentration gradient. 
It is shown in Sect.~\ref{sec:gaussianproc} that 
the velocity field diverges at some distance from the membrane
in the normal mode of $\Omega_*^2<1$.
 This divergence can be regarded as representing the resonance of the harmonic oscillator
at that distance.   For a mode showing this local resonance, as discussed in Sect.~\ref{sec:gaussianproc},
we need a cutoff length to calculate the induced-mass density.
As $\Omega_*^2(<1)$ becomes smaller, the local resonance occurs farther from the membrane and
the calculation result becomes less sensitive to the cutoff length.
The cutoff length is not used in
Fig.~\ref{fig:zoku4}, where only modes with $\Omega_*^2>1$ are considered.
For a mode with $\Omega_*^2$ immediately above unity, oscillation close to the
resonance occurs in the mixture, which increases the induced-mass density
as shown in Sect.~\ref{sec:gaussianproc}.
This causes the rapid drop for $K< 0.09$ in Fig.~\ref{fig:zoku4}.  

\subsection{Case of three intersections\label{sec:gaussianthree}}
We have three intersections in Fig.~\ref{fig:cald}(b), where $I_1$ has $\Omega_*^2\ll 1$ and
the other two have $\Omega_*^2\approx 1$.
In this case, the fraction $1/\left( \rho^{({\rm eff})}_k \omega_*^2 \right)$ on the rhs of Eq.~(\ref{eqn:equip})
is replaced by the sum of the fractions for the three modes.  
As compared with the term from $I_1$, however, the sum of 
the two terms from $I_2$ and $I_3$ is negligible because
$I_1$ gives much smaller values of $\rho^{({\rm eff})}_k$ and $\omega_*^2$
than the latter two modes.  
Thus, when $I_1$ has $\Omega_*^2\ll 1$, regarding the modes of $I_2$ and $I_3$ 
as immobile, we can calculate the MSA by using Eq.~(\ref{eqn:equip}).
Results obtained by this procedure
are graphically shown in Fig.~\ref{fig:dC7h7bun}(b) later. 
Here, we derive an approximate formula for them. 
For $\Omega_*^2\ll 1$, we find that
the induced-mass density is approximately equal to $2\rho/k$, as
mentioned at the end of Sect.~\ref{sec:beyond} later, and we
can use $d_1\approx -1/2$ in
Eqs.~(\ref{eqn:equip}) and (\ref{eqn:perpendrep}), judging from Fig.~\ref{fig:cald}(b).  Thus, 
for a tensionless membrane, we obtain
\begin{equation}
\langle {\hat \zeta}(\bm{ k},t){\hat \zeta}(\bm{ k}',t)\rangle
\approx \delta_{\bm{ k},-\bm{ k}'}{k_{\rm B}T_{\rm c}\over l_{\rm p}^2}
\left[{c_{\rm b}k^4\over 2} + {h^2 \over M\xi_{\rm c}}  {K^2\over K+1} \right]^{-1}
\label{eqn:mainhalf}
\end{equation} 
for $K^3\ll 4\lambda^2$, {\it i.e.\/}, $k^3\ll 4h^2/(c_{\rm b}M)$. 
Comparing the second term in the braces above with that of 
Eq.~(\ref{eqn:mainato}), we find that the numerical factor of 
the restoring force due to the ambient near-criticality for small $K^3/\lambda^2$ is 
reduced to approximately half the factor for large $K^3/\lambda^2$.
The second term, regarded as $h^2k^2\xi_{\rm c}/M$ because $K\ll 1$, 
increases with $\xi_{\rm c}$, which represents the thickness of the adsorption layer.
When $k$ is small enough to
satisfy $K^2 < 2\lambda^2$ furthermore, 
the second term ($\propto k^2$) becomes larger than the first term  ($\propto k^4$) 
in the braces of Eq.~(\ref{eqn:mainhalf}) and
the suppression effect due to the ambient near-criticality becomes distinct.
\\

In the appendix of Ref.~\onlinecite{pre}, the present author tentatively neglected the hydrodynamic effects to
calculate the MSA by using Eq.~(\ref{eqn:glw}) naively.
The result is Eq.~(\ref{eqn:main}) with $K^2/(K+1)$ replaced by $1-1/\sqrt{1+K^2}$, which
is approximately equal to $K^2/2$ for $K\ll 1$.
Thus, this naive calculation also leads to
the reduction in the restoring force by approximately half, which means that the reduction
is caused by the decreased contribution from the hydrodynamics.  This decrease  
for small $k$ is reasonable, considering that
the transient profiles of the fields around a
membrane undulating slowly would be close to their respective equilibrium ones around a membrane whose shape
is fixed at each time.  

\subsection{Calculation procedure \label{sec:gaussianproc}}
We here suppose $h\ne 0$ for simplicity of the description;
$Z\equiv z/\xi_{\rm c}$ is assumed to be positive.
We use the same dimensionless fields as  given by
Eq.~(45) of Ref.~\onlinecite{pre}, {\it i.e.\/},
\begin{equation}
U\left(\bm{k}, Z, \omega\right)\equiv \displaystyle{{i{\tilde V}_z^{(1)}\left(\bm{k},z,\omega\right)
\over \omega {\tilde \zeta}^{(1)}\left(\bm{k},\omega\right)}} 
\ ,\label{eqn:defU}\end{equation}
 which gives the definition of $U_{\rm out}$ if ${\tilde V}_z^{(1)}$ is replaced by
${\tilde V}_{{\rm out}\ z}^{(1)}$.
Hereafter, 
$\partial_Z$ implies the partial derivative with respect to $Z$ and
$\partial^2_Z$ denotes $\partial_Z\partial_Z$. 
The closed equation mentioned below Eq.~(\ref{eqn:newouter3}) is
\begin{equation}
\left(1-\Omega^{2}e^{2Z} \right)(\partial_Z^2-K^2)U_{\rm out}(Z)
=2\partial_Z U_{\rm out}(Z)\ ,
\label{eqn:kikamoto}\end{equation}
where we abbreviate $U_{\rm out}(\bm{k}, Z, \omega)$ as $U_{\rm out}(Z)$.
The boundary conditions are given by $U_{\rm out}(Z)\to 0$ as $Z\to \infty$ and $U_{\rm out}(0+)=1$.
The latter is derived from Eqs.~(\ref{eqn:vzl}) and (\ref{eqn:Voutlim}), and
can also be derived from Eq.~(\ref{eqn:Usol1}).   
Let us introduce ${\cal U}_{\rm o}(\theta)$ so that 
$U_{\rm out}(\bm{k}, Z,\omega)=e^{(1-K_1)Z}{\cal U}_{\rm o}(\theta)$ holds, 
where we use $\theta\equiv \Omega^{2}e^{2Z}$ and
$K_1\equiv \sqrt{1+K^2}$. 
As discussed in Appendix \ref{app:a}, ${\cal U}_{\rm o}$ satisfies Gauss' differential equation. 
Defining $a\equiv \left( 1-K-K_1\right)/2$ and 
$b\equiv \left( 1+K-K_1\right)/2$, we find that
${\cal U}_{\rm o}(\theta) = g(\theta)/g(\Omega^{2})$, 
where $g$ is defined as
\begin{equation}
g(\theta)\equiv \left\{ \begin{array}{ll}
\theta^{-b} {\cal F}\left(1-a, b, 1+b-a, \theta^{-1}\right) & {\rm for}\ \theta>1\\
\left[ f_0^+(\theta) + f_0^-(\theta)\right] / 2 & {\rm for}\ 0<\theta<1
\end{array}\right.
\ .\label{eqn:finf}\end{equation}
Here, ${\cal F}$ denotes Gauss' hypergeometric series, while
each of $f_0^\pm$ is a linear combination of hypergeometric series defined by Eq.~(\ref{eqn:fopfom}); 
$g(\theta)$ diverges logarithmically at $\theta=1$.  If $\Omega^2$ is smaller than unity,
a positive value of $Z$ 
gives $\theta\equiv \Omega^2e^{2Z}=1$.  With 
$Z^*$ denoting this value,  $U_{\rm out}(Z)$ then diverges logarithmically at $Z=Z^*$. 
No divergence occurs otherwise.  \\

\begin{figure}
\includegraphics[width=8cm]{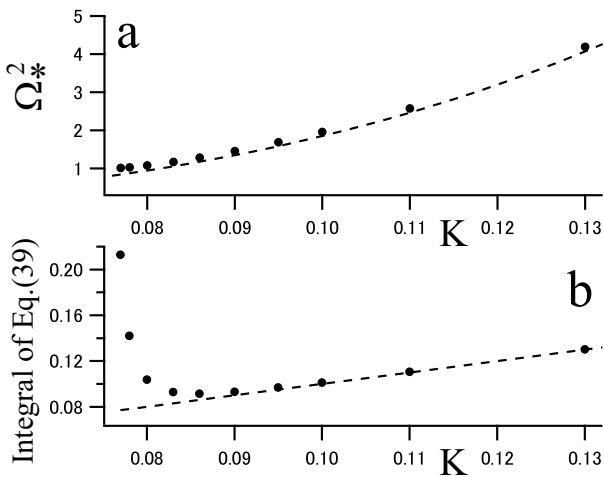}
\caption{\label{fig:zoku4efom} 
Graphical representations of the normal-mode frequency and induced-mass density.  
The material constants are the same as used in Fig.~\ref{fig:zoku4}.
(a) Values of $\Omega_*^2$ are plotted against $K$ (circles).
The dashed line represents $K^3/(4\lambda^2)$.  
(b) Circles represent values of the integral of Eq.~(\ref{eqn:effective}), while
the dashed line represent the value of $K$.   }
\end{figure}

In each normal mode, 
${\hat V}^{(1)}_{{\rm out} \ z}(\bm{k}, z, t)$ equals $U_{\rm out}(\bm{k}, Z, \omega_*)
\partial {\hat \zeta}^{(1)}(\bm{k}, z,t)/(\partial t)$
because of Eq.~(\ref{eqn:defU}).
Thus, Eqs.~(\ref{eqn:kineticenergy}) and (\ref{eqn:kinetic}) give
\begin{eqnarray}&&
\rho^{({\rm eff})}_k=\rho_{\rm m}^{(0)}+{2\rho\over k}\nonumber\\&&
+{2\rho\over kK}
\left\{ -K+\int_0^\infty dZ\ 
\left[\left(\partial_z U_{\rm out} \right)^2 +K^2 U_{\rm out}^2\right]\right\}
\ .\label{eqn:effective} \end{eqnarray}
We obtain the dimensionless MSA in Fig.~\ref{fig:zoku4} by using
Eq.~(\ref{eqn:equip}), {\it i.e.\/}, by adding the sum in the braces of Eq.~(\ref{eqn:perpendrep}) to 
$2\Omega_*^2\lambda^2$ multiplied by the sum in the brackets of  Eq.~(\ref{eqn:effective}).
Figure \ref{fig:zoku4efom} shows the values used in obtaining the circles in Fig.~\ref{fig:zoku4}.  
In Fig.~\ref{fig:zoku4efom}(a), $\Omega_*^2$ is shown to
agree with $K^3/4\lambda^2$, as mentioned
in the last paragraph of Sect.~\ref{sec:gaussiannormal}.  In Fig.~\ref{fig:zoku4efom}(b),
the integral of Eq.~(\ref{eqn:effective}) is shown to
agree with $K$ for large $K$, which is reasonable because
$U_{\rm out}\approx e^{-KZ}$ for $\Omega^2\gg 1$ from Eq.~(\ref{eqn:Usol1}). 
Thus, we have
$\rho^{({\rm eff})}_k\approx \rho_{\rm m}^{(0)}+2\rho/k\approx 2\rho/k$
for a mode with $\Omega_*^2\gg 1$, as mentioned above Eq.~(\ref{eqn:main}).
Figure \ref{fig:zoku4efom}(b) also shows that the induced-mass density increases 
rapidly for $K\stackrel{<}{\sim} 0.09$,
which mainly causes the rapid drop in Fig.~\ref{fig:zoku4}.  \\

For a normal mode with $\Omega_*^2<1$, 
the integrand in Eq.~(\ref{eqn:effective}) 
is almost proportional to $(Z-Z^*)^{-2}$ near $Z=Z^*$.
We replace the integrand for $\left|Z-Z^*\right|<0.05$ with
the Cauchy--Lorentz distribution whose full width at half maximum
is $0.1$. This cutoff length for $Z\equiv z/\xi_{\rm c}$ is introduced considering that the Gaussian model cannot describe 
phenomena with a length scale much smaller than $\xi_{\rm c}$.  
As $\Omega_*^2(<1)$ decreases, $Z^*$ becomes larger and
the integral of the Cauchy--Lorenz distribution over the region of $\left|Z-Z^*\right|<0.05$
becomes much smaller than the whole integral, as shown at the end of Sect.~\ref{sec:beyond} later.
Then, the whole integral is rather insensitive to the cutoff length.  
When $\Omega_*^2$ is sufficiently small, 
$\theta$ remains close to zero and ${\cal U}_{\rm o}(\theta)$ remains close to unity
up to a sufficiently large value of $Z$.   In this region of $Z$, 
we have $U_{\rm out}(\bm{k}, Z,\omega)\approx e^{(1-K_1)Z}$, and have $\mu^{(1)}\approx 0$
because of Eqs.~(\ref{eqn:newouter}) and (\ref{eqn:newouter3}).  This
means that $\mu$ is constant in large mixture regions on both sides around the membrane,
which explains why we
can neglect the hydrodynamic effects for $K^3\ll 4\lambda^2$.

\section{General Model for the Critical Composition \label{sec:general}}
In Eq.~(\ref{eqn:mainhalf}), the suppression effect due to the ambient near-criticality
increases with the correlation length.  To examine whether this tendency continues
beyond the regime of the Gaussian model, we use a more general model introduced in  
the renormalized local functional theory \cite{JCP, PRE, yabu}.   
We assume that $\varphi_\infty$ equals the critical value of $\varphi$, and
that the reduced temperature, $\tau\equiv \left(T-T_{\rm c}\right)/T_{\rm c}$, is positive; 
$T_{\rm c}$ denotes the critical temperature.  The critical exponents
$\beta\approx 0.326$, $\gamma\approx 1.239$, $\eta\approx 0.024$, and $\nu\approx 0.627$
are the same as the ones for the three-dimensional Ising model \cite{Onukibook}. Here, 
$\beta$ and $\gamma$ respectively involve the singularities of the order parameter and susceptibility,
while $\eta$ involves the power decay of the correlation function at the critical point.  
The correlation length far from the membrane, denoted by
 $\xi_\infty$, becomes $\xi_0\tau^{-\nu}$ in the leading behavior as $\tau\to 0+$,
where $\xi_0$ is a nonuniversal constant.  \\

The free-energy functional is shown in Sect.~\ref{sec:genfree}, while the results are shown
in Sect.~\ref{sec:beyond}, where
we focus on the modes each of which has so small wavenumber  
that the hydrodynamic effects of the mixture are negligible.  
A study on the hydrodynamic effects 
beyond the regime of the Gaussian model, given in Appendix \ref{app:hydgen}, suggests
that Fig.~\ref{fig:cald} remains almost unchanged beyond the regime if $\xi_{\rm c}$ is replaced by 
$\xi(0+)$, which is defined as the correlation length very near the membrane in the unperturbed state.  
Thus, the condition under which we can neglect the hydrodynamic effects of the mixture
will remain  $k^3\ll 4h^2/(c_{\rm b}M)$ beyond the regime. 
This can be expected because this inequality is independent of the correlation length. 
We also assume $k\xi_\infty \ll 1$ in Sect.~\ref{sec:beyond}. 

\subsection{Free-energy functional and the unperturbed profile\label{sec:genfree}}
The correlation length $\xi$
can be inhomogeneous; $w\equiv \xi_0^{1/\nu}\xi^{-1/\nu}$
represents a local ``distance'' from the critical point and depends not only on $\tau$
but also on the local value of $\varphi$. The dependence is given by Eq.~(\ref{eqn:wtaupsi}) below. 
In the renormalized local functional theory, $M(\varphi)$
of Eq.~(\ref{eqn:glw}) is given by
\begin{equation}
M(\varphi) \equiv k_{\rm B}T_{\rm c} C_1w^{-\eta\nu}
\ ,\label{eqn:Cdef}\end{equation}
where $C_1$ is a nonuniversal constant, while $f(\varphi)$ is replaced by
\begin{equation}
 f_{\rm R}(\psi)\equiv {k_{\rm B}T_{\rm c}\over 4} \left( 2 C_1\xi_0^{-2}w^{\gamma-1}\tau\psi^2 
 +C_1^2 u^*\xi_0^{-\epsilon}w^{\gamma-2\beta}\psi^4\right)
\label{eqn:f}\end{equation}
as a result of the $\epsilon$-expansion.   Here,
$\psi\equiv \varphi-\varphi_\infty$ is the order parameter.  In later numerical calculations,
we use $\epsilon=1$ and approximate the constant $u^*$ as $2\pi^2/9$.  
The scaling and hyperscaling laws \cite{Onukibook}, $\gamma=\nu(2-\eta)$ and
$(4-\epsilon)\nu=2\beta+\gamma$, give $\left(\epsilon-2\eta\right)\nu=\gamma-2\beta$,
and thus Eq.~(\ref{eqn:f}) is the same as Eq.~(3.5) of Ref.~\onlinecite{JCP}. 
We can consider $\mu^{(0)}$ to vanish in the critical composition \cite{com2}.
The critical fluctuation of the mixture
 is noteworthy only on the length scales smaller than the correlation length, while 
Eq.~(\ref{eqn:f}) is renormalized up to the local
correlation length. Thus, we can use the mean-field theory 
to calculate $\xi$ at each locus; this leads to
\begin{equation}
w=\tau+C_2w^{1-2\beta}\psi^2
\ ,\label{eqn:wtaupsi}\end{equation}   
where $C_2\equiv 3 u^* C_1 \xi_0^{2-\epsilon}$ is a nonuniversal constant.
These equations can be found in Ref.~\onlinecite{JCP}; see its
Eqs.~(3.9) and (3.11) and the statement below its Eq.~(3.16).  
When $\tau$ is not too small, 
large values of $|\psi|$ are infrequent and the term involving $\psi^4$
is negligible in Eq.~(\ref{eqn:f}) \cite{later}. 
Then, having $w^{-\eta\nu}\approx \tau^{-0.015}\approx 1$, we can regard  
Eq.~(\ref{eqn:f}) as 
$k_{\rm B}T_{\rm c} C_1\psi^2/(2\xi_\infty^2)$, which represents the Gaussian model
with $k_{\rm B}T_{\rm c} C_1$ and $\xi_\infty$ being regarded respectively as $M$ and $\xi_{\rm c}$.  \\

We use the unperturbed profile $\varphi^{(0)}$ in our perturbative calculation; Eq.~(\ref{eqn:phizero}) is valid 
only in the regime of the Gaussian model.
As discussed in Sect.~IIB of Ref.~\onlinecite{JCP}, even when the mixture lies beyond the regime of the Gaussian 
model, we can still obtain $\varphi^{(0)}$ by minimizing the free-energy functional,  
Eq.~(\ref{eqn:glw}) with Eqs.~(\ref{eqn:Cdef}) and ({\ref{eqn:f}), in the renormalized local functional theory \cite{com2}.
This is consistent with the use of the mean-field theory in deriving Eq.~(\ref{eqn:wtaupsi}). 
The unperturbed profile satisfies
\begin{equation}
 M(\varphi(z))\left| {d\psi(z)\over dz}\right|^2= 2 f_{\rm R}(\psi(z)) 
\label{eqn:bulk}\end{equation}
and \begin{equation}
M(\varphi(z)) {d\psi(z)\over dz}=-h\quad {\rm as}\ z\to 0+\ .
\label{eqn:bc}\end{equation}
In the unperturbed state, the membrane can be regarded as a wall
with the preferential attraction.  In a mixture having the critical composition
far from the wall,  the second-order transition at $T= T_{\rm c}$ 
is called the normal transition, where  
the order-parameter profile causes  
the critical adsorption \cite{liu,diehl94}. 
The same universality is shared by
the extraordinary transition, which occurs  
in the three-dimensional Ising model in a finite lattice in the absence of bulk and surface 
fields \cite{bray, binder, diehl86, burk, diehl97}.
In the usual renormalization group theory, 
the order parameter is coarse-grained and rescaled, and
the surface field appears to diverge at the normal transition \cite{diehl97,smock,enhan}.
However, Eq.~(\ref{eqn:f}) has
the order parameter on its original scale, and thus we can use a finite surface field on its original scale
to obtain Eq.~(\ref{eqn:bc}).
As shown in Appendix A of Ref.~\onlinecite{sphere}  and in Appendix \ref{app:procgen} of the present study,  
Eqs.~(\ref{eqn:bulk}) and (\ref{eqn:bc}) generate the profile 
obtained in the previous works.
More details are shown in Appendix \ref{app:unpgen}. 
\\

As mentioned in Sect.~\ref{sec:intro}, 
we can suppose a membrane immersed in some organic mixture.  
In our numerical calculations, we 
use $T_{\rm c}=300$ K and $\xi_0=0.23$ nm, which are data for a mixture
of nitroethane and 3-methylpentane \cite{iwan}.   
A membrane with outward nonpolar chains
has $c_{\rm b}=2 \times 10^{-20}$ J $=4.2 k_{\rm B}T_{\rm c}$ \cite{lei2}. 
On the basis of the discussion in Ref.~\onlinecite{liu}, 
we use $h= 10^{-7}$ m$^3$/s$^2$ \cite{relax}.
As discussed in Ref.~\onlinecite{pre}, we use
$M=k_{\rm B}T_{\rm c}C_1=10^{-16}$ m$^7$/(s$^2$kg) \cite{aiche, kahl, vdwexp}.
In Fig.~\ref{fig:zeh7}, $\xi(0+)$
reaches a plateau as $\tau$ decreases, unlike $\xi_\infty$.
This is reasonable, considering that the surface field prevents
the mixture near the membrane from approaching the critical composition. 
As discussed below Eq.~(\ref{eqn:gaussian}),
$\xi(0+)$ agrees with $\xi_\infty$ in the regime of the Gaussian model,
where $\xi_\infty$ is given by $\xi_{\rm c}$.
 In this regime, we have
$\tau \stackrel{>}{\sim} 3\times 10^{-3}$ and $\xi_{\rm c}\stackrel{<}{\sim} 10$ nm in Fig.~\ref{fig:zeh7}.  

\begin{figure}
\includegraphics[width=8cm]{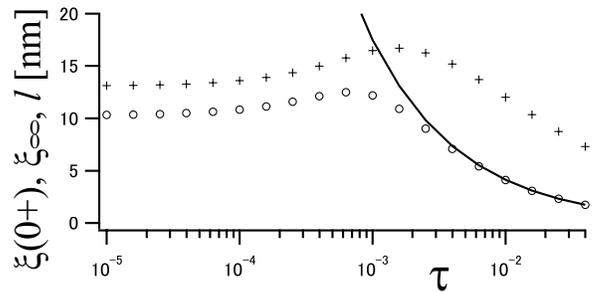}
\caption{\label{fig:zeh7} Numerical results of $\xi(0+)$ (circles), which are obtained 
by using Eq.~(\ref{eqn:xi0+}).  
Crosses represent those of $l$, whose definition and calculation procedure are shown
in Appendix \ref{app:procgen}.     
The solid curve represents $\xi_\infty=\xi_0\tau^{-\nu}$, which
reaches $7.4\times 10$ nm at $\tau=10^{-4}$.
The material constants used are mentioned in Sect.~\ref{sec:genfree}.}
\end{figure}

\subsection{Results in the general model\label{sec:beyond}}
We apply 
the naive way, mentioned at the end of Sect.~\ref{sec:gaussianthree},
to calculate the MSA  for the modes of $k^3\ll 4h^2/(c_{\rm b}M)$ and $k\xi_\infty\ll 1$, which are
free from distinct hydrodynamic effects and critical fluctuation. 
After the calculation 
shown in Appendix \ref{sec:naive} for a 
tensionless membrane, we arrive at
\begin{equation}
\langle {\hat \zeta}(\bm{ k},t){\hat \zeta}(\bm{ k}',t)\rangle
\approx \delta_{\bm{ k},-\bm{ k}'}{k_{\rm B}T_{\rm c} \over l_{\rm p}^2}
\left[ {c_{\rm b}k^4\over 2} + D(k)
\right]^{-1}\ .\label{eqn:main3}
\end{equation} 
Here, we write $D(k)$ for the term due to the ambient near-criticality.  It is given by
\begin{eqnarray}
&&D(k)= 2k^2\int_0^\infty dz\ M(\varphi^{(0)}(z))\left({d \varphi^{(0)}\over d z}\right)^2 \label{eqn:Ddef1}\\
&&\qquad =4k^2\int_0^\infty dz\ f_{\rm R}(\psi^{(0)}(z))\ ,
\label{eqn:Ddef2}\end{eqnarray}
where $\psi^{(0)}(z)$ denotes the unperturbed profile of the order parameter, {\it i.e.\/}, $\varphi^{(0)}(z)-\varphi_\infty$.
The second equality comes from Eq.~(\ref{eqn:bulk}); 
$D(k)$ is proportional to $k^2$.  
Substituting Eq.~(\ref{eqn:phizero})
into Eq.~(\ref{eqn:Ddef1}) gives $D(k)=h^2k^2\xi_{\rm c}/ M$.
This means that  Eq.~(\ref{eqn:main3}) becomes
Eq.~(\ref{eqn:mainhalf}) in the regime of the Gaussian model.
For sufficiently small $\tau$ beyond the regime,
assuming nonzero $h$ and $k\xi(0+)\ll 1$, 
we use $u^*=2\pi^2/9$ to obtain 
\begin{eqnarray}
&&D(k)\approx {\sqrt{6}\beta\over 4\nu \pi^2 }\times
 { k_{\rm B}T_{\rm c} k^2   \over  \xi(0+)^2} 
\label{eqn:Dapp2} \\ &&\qquad
\approx {\sqrt{6}\beta\over \nu}\times {h^2k^2\xi(0+) \over M_{00}}
\ ,\label{eqn:Dapp2ato}\end{eqnarray}
where $M_{00}$ denotes $M(\varphi^{(0)}(0+))$.
Equation (\ref{eqn:Dappapp}) leads to Eq.~(\ref{eqn:Dapp2}), while 
Eq.~(\ref{eqn:tuika}) helps in deriving Eq.~(\ref{eqn:Dapp2ato}). 
Judging from Fig.~\ref{fig:zeh7}, 
$M_{00}$ remains almost the same as $k_{\rm B}T_{\rm c}C_1$  
because $\left[\xi(0+)/\xi_0\right]^\eta \approx 1$.   
Because $\sqrt{6}\beta/\nu\approx 1.3$, we can
obtain Eq.~(\ref{eqn:Dapp2ato}) roughly by replacing $\xi_{\rm c}$ with $\xi(0+)$ in the second term
in the braces of Eq.~(\ref{eqn:mainhalf}).   As shown in Fig.~\ref{fig:zeh7}, 
$\xi(0+)$ reaches a plateau as $\tau$ decreases.  
Hence, the term due to the ambient near-criticality increases with $\xi_\infty$ 
only in the regime of the Gaussian model.
\\

\begin{figure*}
\includegraphics[width=14cm]{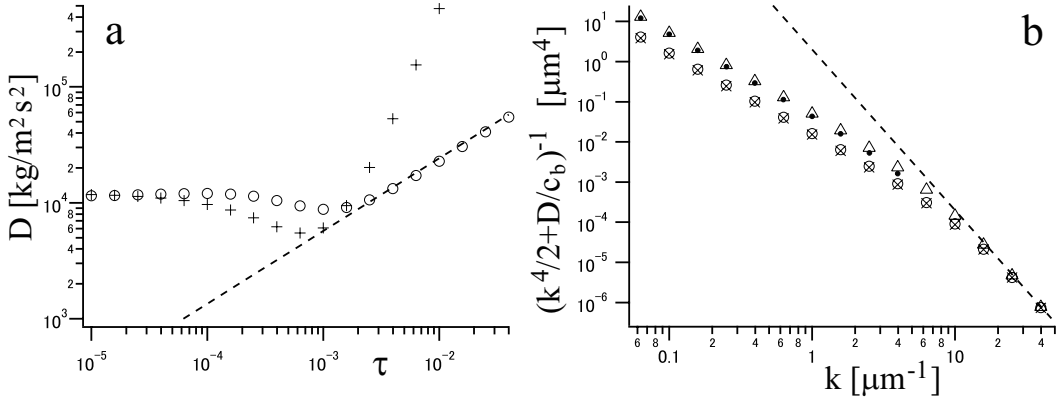}
\caption{\label{fig:dC7h7bun} 
 (a) Numerical results of $D$.
Circles and crosses represent Eqs.~(\ref{eqn:Ddef2}) and
(\ref{eqn:Dapp2}) for $k\xi(0+)=10^{-3}$, respectively.
The dashed line represents $(hk)^2\xi_{\rm c}/\left[ M\left(k\xi_{\rm c}+1\right)\right]$ with
$\xi_{\rm c}$ put equal to $\xi_\infty=\xi_0 \tau^{-\nu}$.
(b) Graphical representation of the MSA for sufficiently small $k$.
 For $\tau=10^{-5}$ $(\circ)$, $10^{-3}$ $(\times)$, and $10^{-2}$ $(\triangle)$,  we 
use  Eq.~(\ref{eqn:Ddef2}) to plot the reciprocal of $(k^4/2)+(D/c_{\rm b})$, which is 
a part of the rhs of Eq.~(\ref{eqn:main3}). 
The values of $\xi_\infty$ are $3.1\times 10^{-1}$, $1.7\times 10^{-2}$, and $4.1\times 10^{-3}\ \mu$m, respectively,
for these values of $\tau$.
Solid circles represent the results of the Gaussian model with $\xi_{\rm c}$ put equal to
$\xi_\infty$ at $\tau=10^{-2}$ and with $\lambda=1.9\times 10^{-2}$;  
$\Omega_*^2$ equals $1.1\times 10^{-2}$ on the extreme right.
The dashed line represents $2/k^4$.
The material constants mentioned 
in Sect.~\ref{sec:genfree} are used in (a) and (b). }
\end{figure*}

These findings are shown in Fig.~\ref{fig:dC7h7bun}. 
We numerically calculate
Eq.~(\ref{eqn:Ddef2}) in the way mentioned in the last paragraph of Appendix \ref{app:procgen} and find that 
$D(k)$ for $k\xi(0+)=10^{-3}$ reaches a plateau when $\tau$ takes values
from $10^{-5}$ to $10^{-3}$ in Fig.~\ref{fig:dC7h7bun}(a).
The plateau value agrees with Eq.~(\ref{eqn:Dapp2}).
In this range of $\tau$, $\xi(0+)$ also reaches a plateau in Fig.~\ref{fig:zeh7}. 
For $\tau \stackrel{>}{\sim} 10^{-3}$ in Fig.~\ref{fig:dC7h7bun}(a), the circles agree well with
the dashed line representing the corresponding term of Eq.~(\ref{eqn:mainhalf}) obtained in the Gaussian model,
where $\xi_{\rm c}=\xi_\infty$ is identified with $\xi(0+)$. 
The vertical axis in Fig.~\ref{fig:dC7h7bun}(b), as labeled, represents a  part of the rhs of
Eq.~(\ref{eqn:main3}).  Although data are not shown, the results
for  $\tau=10^{-4}$ are 
almost the same as the ones for $\tau =10^{-3}$, as expected from
Fig.~\ref{fig:dC7h7bun}(a).
In Fig.~\ref{fig:dC7h7bun}(b), because of $4h^2/(c_{\rm b}M)=(3\times 10\ \mu{\rm m}^{-1})^3$,
the range of $k\stackrel{<}{\sim} 10 \ \mu$m$^{-1}$ meets the validity condition
$k^3\ll 4h^2/(c_{\rm b}M)$.
We also require $k\xi_\infty\ll 1$, which lowers the upper limit of the range of validity 
only for $\tau<10^{-4}$.   For small $k$ in Fig.~\ref{fig:dC7h7bun}(b),
the results distinctly deviate below the dashed line, which means that the second term
is predominant in the braces of Eq.~(\ref{eqn:main3}).
Thus, for the material constants used,
the suppression effect due to the ambient near-criticality should be experimentally detectable 
when the wavelength is longer than some micrometers and much longer than $\xi_\infty$. 
\\

The triangles in Fig.~\ref{fig:dC7h7bun}(b) can be regarded as representing the results in the Gaussian model
without the hydrodynamic effects being considered because, as shown in Fig.~\ref{fig:dC7h7bun}(a), 
$\tau=10^{-2}$ is well in the regime of the Gaussian model.
In Sect.~\ref{sec:gauss},
we take into account hydrodynamic effects 
in the Gaussian model.  As discussed in Sect.~\ref{sec:gaussianthree},  
three normal modes can exist for a wavenumber vector $\bm{k}$; 
we can regard the normal modes of $\Omega_*^2\approx 1$ as immobile and 
use the cutoff length for the other mode if $k$ is sufficiently small.
This procedure, being valid for $k<5\ \mu$m$^{-1}$,  
yields the solid circles in Fig.~\ref{fig:dC7h7bun}(b).  For them,
the ratio of the integral of the Cauchy--Lorentz distribution, mentioned in
the last paragraph of Sect.~\ref{sec:gaussianproc}, 
decreases with $k$ from $20$\% to $1$\%.  
For small $k$, the solid circles agree with the triangles, and thus
Eq.~(\ref{eqn:mainhalf}) gives a good approximation  
in the Gaussian model.  
The quotient of the difference in the brackets of Eq.~(\ref{eqn:effective}) divided by $K$ 
is almost proportional to $k$ for the solid circles and is 
$1.1\times 10^{-1}$ at $k= 1\ \mu$m$^{-1}$.  Below this value of $k$, we
can regard  $\rho^{({\rm eff})}_k$ as $2\rho/k$, as mentioned above Eq.~(\ref{eqn:mainhalf}).
For each of the circles, the major contribution to the integral of Eq.~(\ref{eqn:effective}) comes 
from the region $Z>Z^*+0.05$.
Because $g(\theta)\approx \theta^{-b}$ for $\theta \gg 1$ and  $g(\Omega^2)\approx 1$ for
$\Omega^2\ll 1$ and $K\ll 1$,  
we have $U_{\rm out}(\bm{k}, Z, \omega_*)\approx e^{-KZ}/\Omega_*^{2b}$ for $Z\stackrel{>}{\sim} Z^*$,
which makes the quotient almost vanish for sufficiently small $k$.
For $10\ \mu$m$^{-1}\stackrel{<}{\sim}k\stackrel{<}{\sim} 20\ \mu$m$^{-1}$ in Fig.~\ref{fig:dC7h7bun}(b), 
three normal modes exist for a wavenumber vector $\bm{k}$, and
the results become smaller than half the values shown by the dashed line
because the local resonance significantly increases the induced-mass density. 
However, they are rather sensitive to the cutoff length and not shown in the figure.

\section{\label{sec:dis} Summary and Outlook}
By performing calculations within the linear approximation with respect to $\zeta$, we here study
how the ambient near-criticality, combined with the preferential attraction, influences
$\langle {\hat \zeta}(\bm{k},t) {\hat \zeta}(\bm{k}',t)\rangle$ of a tensionless fluid membrane.  
Clearly, we should take into account possible effects due to the reversible dynamics of the ambient mixture,  
which moves with the membrane.  
Assuming $k\xi_{\rm c}\ll 1$, we derive
a closed equation for ${\tilde V}^{(1)}_{{\rm out}\ z}$, devise a procedure for relating its solution
with the stress exerted on the membrane, and calculate the MSA in the Gaussian model.  
The dependence of the MSA on $k$ is elucidated as follows.
Suppose that $k$ decreases from a large value making the ratio $c_{\rm b}Mk^3/(4h^2)$ much larger than unity.
Until the ratio becomes close to unity, the ratio approximately equals $\Omega_*^2$ 
and the previous result of Ref.~\onlinecite{pre} is valid, 
as shown in Figs.~\ref{fig:cald}(a), \ref{fig:zoku4}, and \ref{fig:zoku4efom}(a). 
When $\Omega_*^2$ is larger than and close to unity,
the MSA is markedly suppressed because the induced-mass density increases,
which is shown by the rapid drop in Fig.~\ref{fig:zoku4} and the rapid increase in Fig.~\ref{fig:zoku4efom}(b).
Hydrodynamic effects are negligible 
when the ratio is much smaller than unity, as discussed 
in Sect.~\ref{sec:gaussianthree} and shown 
by the agreement between the triangles and solid circles in Fig.~\ref{fig:dC7h7bun}(b).
For the results of these symbols, the suppression effect due to the ambient near-criticality
becomes distinct and predominant over the one due to the 
bending rigidity.  
In the range $10\ \mu$m$^{-1}\stackrel{<}{\sim} k \stackrel{<}{\sim}20\ \mu$m$^{-1}$
of Fig.~\ref{fig:dC7h7bun}(b), where the ratio lies between approximately $0.05$ and $0.5$,
the MSA is suggested to be suppressed because of the local resonance mentioned in Sect.~\ref{sec:gaussiansingle},
but future works possibly involving nonlinearity are required to calculate reliable results in this range. 
It would be interesting to observe this suppression, as well as the rapid drop mentioned above, experimentally.
The latter would be detectable in a small membrane as far as contained there.
\\

Beyond the regime of the Gaussian model, we use the renormalized local functional theory \cite{JCP}
to calculate the MSA for values of $k$ which are small enough to make the hydrodynamic effects
of the mixture negligible.  
The critical composition is assumed far from the membrane.  As shown in Fig.~\ref{fig:dC7h7bun}(a),
the result turns out to be given roughly by the corresponding result in the Gaussian model 
with $\xi_{\rm c}$ being replaced by $\xi(0+)$, which reaches a plateau as 
 the temperature approaches the critical one (Fig.~\ref{fig:zeh7}).
Thus, as shown in Fig.~\ref{fig:dC7h7bun}(b), the suppression effect due to the ambient near-criticality for small $k$ 
remains distinct, but does not increase with $\xi_\infty$, 
beyond the regime of the Gaussian model.   Some considerations on the hydrodynamic effects in Appendix \ref{app:hydgen}
suggest that the above-mentioned replacement is also valid in a range of larger
values of $k$. 
We assume $k\xi_\infty\ll 1$ in the present formulation; the
distinct concentration fluctuation of the mixture should affect the MSA
without this assumption.   This line of investigation awaits future study.  \\

 It is expected that 
the present study will help in determining the material constants $M$ and $h$ from 
experimental data. 
A sufficiently large membrane can be prevented  from becoming floppy by slightly decreasing the temperature
towards the critical value. 
In considering the hydrodynamic effects of the mixture, we do not assume
its viscosity and diffusion coefficient.  This is not because they are negligible in 
a real mixture but because they do not affect the equal-time correlation, as mentioned in Sect.~\ref{sec:intro}.
In practice, the local resonance in the velocity field of the mixture
 could not be observed in the
shape fluctuation about the equilibrium, considering that
its relaxation is usually overdamped \cite{relax}.\\

The suppression effect due to the ambient near-criticality 
has also been pointed out theoretically for a trapped Brownian spherical particle
in the Gaussian model supposing weak preferential attraction \cite{sphere}, as was done
for  an almost planar membrane in Ref.~\onlinecite{pre}.
The present study on the membrane will also provide solid foundations in extending the study on the sphere.   \\

\begin{acknowledgments}
The author thanks Professor Shun Shimomura for helpful comments on 
the analytical continuation of the hypergeometric function.  This work was 
partly financed by Keio Gakuji Shinko Shikin.
\end{acknowledgments}

\appendix
\section{\label{app:a} Solution of Eq.~(\ref{eqn:kikamoto})}
Using $a$ and $b$ defined above Eq.~(\ref{eqn:finf}),
we can rewrite Eq.~(\ref{eqn:kikamoto}) as
\begin{equation}
\theta \left(1-\theta\right) {\cal U}_{\rm o}''+\left[a+b-(a+b+1)\theta\right]
{\cal U}_{\rm o}'-ab\ {\cal U}_{\rm o}=0\ ,
\label{eqn:gausskika}\end{equation}
which is Gauss' differential equation \cite{erdel, abr}.  
The governing equations and boundary conditions in Sect.~\ref{sec:form} determine 
the mixture fields uniquely.  
Because $V_z^{(1)}$ is real, we use Eq.~(\ref{eqn:defU}) to find that
$U_{\rm out}(\bm{k},  Z, \omega)$ and $U_{\rm out}(-\bm{k}, Z, -\omega)$
are complex conjugates.  
They should be identical, being the unique solution of Eq.~(\ref{eqn:kikamoto}), and thus 
$U_{\rm out}$ is real.
This means that the phase difference between the nondissipative oscillations of ${\hat \zeta}$ and ${\hat V}_{{\rm out}\ z}^{(1)}$
is $\pi$, considering Eq.~(\ref{eqn:defU}). \\

In the complex plane of $\theta$,
Eq.~(\ref{eqn:gausskika}) has regular singular points at $\theta=0$, $1$, and $\infty$.
Around $\theta=0$, the linearly independent solutions are given by
\begin{eqnarray}
&&f_{01}(\theta)\equiv {\cal F}\left(a,b,a+b,\theta\right) \quad {\rm and} \nonumber\\
&&f_{02}(\theta)\equiv \theta^{1-a-b}{\cal F}\left(1-a,1-b,2-a-b,\theta\right)
\ .\label{eqn:f0102}\end{eqnarray}
Each can be analytically continued
to the solution around $\theta=1$ and
to the solution around $\theta=\infty$. Two linearly
independent solutions in the latter region can be taken so that one
has a factor of $\theta^{-a}$ and the other has a factor of $\theta^{-b}$,
as shown by Eq.~(15.3.7) of Ref.~\onlinecite{abr}.  Because of the boundary condition at $\theta \to\infty$
and $a<0<b$, we arrive at the first line of Eq.~(\ref{eqn:finf}).
For this first line, we write $f_\infty(\theta)$, which diverges logarithmically as $\theta \to 1$
as shown by Eq.~(15.3.10) of Ref.~\onlinecite{abr}. 
We have only to look at the real axis of $\theta$ on the positive side, which 
is covered  by the circles of convergence with partial overlapping. \\

When we have $\Omega^2\ge 1$, 
$g(\theta)$ need not be defined for $0<\theta<1$.
Let us consider the case of $\Omega^2< 1$.
The path of the analytical continuation runs through
the upper half (with the positive imaginary part) of the complex $\theta$-plane
or through the lower half (with the negative imaginary part).
For the results obtained by way of
these paths, we respectively write
\begin{eqnarray}
&&f_{0}^+(\theta)=B_1^+ f_{01}(\theta)+B_2^+ f_{02}(\theta)
\ {\rm and}\nonumber\\
&&f_{0}^-(\theta)=B_1^- f_{01}(\theta)+B_2^- f_{02}(\theta)
\label{eqn:fopfom}\end{eqnarray}
for $|\theta|<1$.
Using Eq.~(15.3.7) of Ref.~\onlinecite{abr}, we have
\begin{eqnarray}&&
B_1^\pm\equiv {\Gamma(2-a-b)\Gamma(b-a)\over J\Gamma(1-a)^2}e^{\mp i\pi b}\ {\rm and}\nonumber\\
&&B_2^\pm\equiv {\Gamma(a+b)\Gamma(b-a)\over J\Gamma(b)^2}e^{\pm i\pi a}
\ ,\label{eqn:Bdef}\end{eqnarray}
where $\Gamma$ denotes the gamma function and we use
\begin{eqnarray}
&&J\equiv \Gamma(2-a-b)\Gamma(b-a)\Gamma(a+b)\Gamma(a-b)\nonumber\\
&&\qquad \times \left[ \sin^2{(\pi a)}-\sin^2{(\pi b)}\right]\pi^{-2}
\ .\end{eqnarray}
Because of the logarithmic divergence at $\theta=1$, 
$f_0^+(\theta)$ and
$f_0^-(\theta)$ for $0< \theta<1$ are not real and are complex conjugates. 
The second line of Eq.~(\ref{eqn:finf}) is thus real, as is required.

\section{\label{app:stress} Relation between ${\tilde F}_z^{(1)}$ and $U_{{\rm out}}$ in the Gaussian model}
As in Ref.~\onlinecite{pre}, we define $G\left(\bm{k}, Z, \omega\right)$ and $Q\left(\bm{k}, Z, \omega\right)$ as
\begin{equation}
\displaystyle{{M {\tilde \varphi}^{(1)}\left(\bm{k},z,\omega\right)
\over h{\tilde \zeta}^{(1)}\left(\bm{k},\omega\right)}} \quad{\rm and}\quad 
\displaystyle{{\xi_{\rm c}^2{\tilde \mu}^{(1)}\left(\bm{k},z,\omega\right)
\over h{\tilde \zeta}^{(1)}\left(\bm{k},\omega\right)}}
\ ,\label{eqn:defG}
\end{equation}
respectively.
We sometimes write $U(Z)$ for $U(\bm{k}, Z, \omega)$ of Eq.~(\ref{eqn:defU}),
and also use similar concise notation for $G$, $Q$, and their outer solutions.
Using
\begin{equation} \Xi(Z)\equiv -e^{-Z} \ ,\label{eqn:Xigauss}
\end{equation}
we rewrite Eq.~(\ref{eqn:eqforvz}) as
\begin{equation}
\left( \partial_Z^2-K^2 \right) U(Z)= 
\Omega^{-2} \Xi(Z)Q(Z) \ .\label{eqn:pZKU}
\end{equation}
We have $U(0+)=1$ because of Eq.~(\ref{eqn:vzl}), while
$U(Z)\to 0$ as $Z\to \infty$ as stated below Eq.~(\ref{eqn:perexp}).
As shown by Eq.~(48) of Ref.~\onlinecite{pre}, we
solve Eq.~(\ref{eqn:pZKU}) to obtain
\begin{eqnarray}
&&U(Z)={1\over \Omega^2} \int_0^\infty d Z_1\ \Gamma_K(Z,Z_1)Q(Z_1)\Xi(Z_1)\nonumber\\&&
+\left[1+{ 1\over 2\Omega^2 K} \int_0^\infty d Z_1\ Q(Z_1)\Xi(Z_1)e^{-KZ_1}\right]e^{-KZ}
\ ,\nonumber\\\label{eqn:Usol1}\end{eqnarray}
where the kernel is defined as
$\Gamma_K(Z,Z_1)=
-e^{-K  \left\vert Z-Z_1 \right\vert}/(2K) $.\\

We consider the limit of $L\to 0+$.
Subtracting $Q_{\rm out}$ from $Q$, we  
define $Q_{\rm in}$ as the difference, which cannot be neglected
in the integrals of Eq.~(\ref{eqn:Usol1}).
Assuming that $Z(>0)$ lies outside the boundary layer in Eq.~(\ref{eqn:Usol1}),  
we find that $U_{\rm out}(Z)$ is independent of $Q_{\rm in}$ and obtain 
\begin{equation}
\lim_{Z\to 0+}\partial_Z U_{\rm out}
=-K-{1\over \Omega^2} \int_0^\infty dZ_1\  Q_{\rm out}(Z_1)\Xi(Z_1)e^{-KZ_1}
\ ,\label{eqn:partZUout}\end{equation}
while Eqs.~(\ref{eqn:Usol1}) and (\ref{eqn:partZUout}) yield
\begin{equation}
\lim_{Z\to 0+}\partial_Z U
=\lim_{Z\to 0+}\partial_Z U_{\rm out}
-{\Xi(0+)\over \Omega^2} \int_0^\infty dZ_1\ Q_{\rm in}(Z_1)\ .\label{eqn:partU}
\end{equation}
The limit of $L\to 0+$ of  Eq.~(\ref{eqn:mu1limitkantan}) gives
\begin{equation}
\lim_{Z\to 0+}\left(\partial_Z G_{\rm out}-
\partial_Z G \right)
=-\int_0^\infty dZ\ Q_{\rm in}(Z)
\ ,\label{eqn:partGout}\end{equation}
where we note that 
$Q_{\rm in}(Z)$ vanishes outside the boundary layer, 
although its integral from $Z=0$ to a positive value 
remains nonzero.  
We can use $\lim_{Z\to 0\pm}\partial_Z G=-1$ because of
Eq.~(\ref{eqn:bcphi}).
\\

Equation (\ref{eqn:newouter}) is nondimensionalized as $G_{\rm out}=-\Xi U_{\rm out}$, which gives
\begin{equation}
\lim_{Z\to 0+}\partial_Z G_{\rm out}=-\Xi(0+) \lim_{Z\to 0+}\partial_Z U_{\rm out}-1
\label{eqn:partGout2}\end{equation}
because $\Xi'(0+)=1$.  Combining
 Eq.~(\ref{eqn:partGout}) with Eq.~(\ref{eqn:partGout2}),
we can delete the integral of $Q_{\rm in}$ from Eq.~(\ref{eqn:partU}) to obtain 
\begin{equation}
 \lim_{Z\to 0+}\partial_Z U
=\left[1-\Omega^{-2}\Xi(0+)^2\right] \lim_{Z\to 0+}\partial_Z U_{\rm out}
\ .\label{eqn:partUU}\end{equation}
We can use Eq.~(\ref{eqn:Xigauss}) to have $\Xi(0+)^2=1$ here. 
Substituting Eq.~(\ref{eqn:partUU}) into Eq.~(\ref{eqn:tilFz}), 
we can calculate ${\tilde F}_z^{(1)}$ by using not $U$
but $U_{{\rm out}}$.
The definitions of Eqs.~(\ref{eqn:defG}) and (\ref{eqn:Xigauss}) are
 generalized in Appendix \ref{app:hydgen}, where
 $\Xi(0+)^2$ is not always unity. \\

The relation between $U_{\rm out}$ and $g$ gives
\begin{equation}
\lim_{Z\to 0+}\partial_Z U_{\rm out}+K =
2\left[ b+\Omega^2 { g'\left(\Omega^{2}\right)\over
g \left(\Omega^{2}\right)} \right]
\ .\label{eqn:gbarg}\end{equation}
We introduce $d_1$ in Eq.~(\ref{eqn:perpendrep}), which gives
\begin{equation}
{(1+d_1)K^2\over K+1}=
\Omega^2 \left( \lim_{Z\to 0+} \partial_ZU  +K \right)
\label{eqn:d1tochu}\end{equation}
because of 
Eqs.~(\ref{eqn:maku}),  (\ref{eqn:tilFz}), and (\ref{eqn:Lamgauss}).
Thus, by using Eqs.~(\ref{eqn:Xigauss}) and (\ref{eqn:partUU}), we obtain
\begin{equation}
d_1={1\over K}+ {2\left(\Omega^{2}-1\right) \left(K+1\right) \over K^2 }
\left[ b+ \Omega^2 { g'\left(\Omega^{2}\right)\over
g \left(\Omega^{2}\right)} \right]
\label{eqn:deeone}\ .\end{equation}
Because Eq.~(\ref{eqn:gbarg}) is real,  
the sum in the braces of Eq.~(\ref{eqn:perpendrep}) is real, as it should be.
The product of Eq.~(\ref{eqn:gbarg}) and  $\Omega^{2}-1$ tends to $0+$ as $\Omega^2$ approaches unity.
Thus, $d_1$ is found to equal $1/K$ at $\Omega^2=1$ from Eq.~(\ref{eqn:deeone}).
In our numerical calculation, $d_1$ increases steeply as $\Omega^2$ approaches
unity but  this exact peak value cannot be obtained. 
Each curve of $d_1$ in Fig.~\ref{fig:cald} is drawn by
connecting the exact peak value and many other numerical outputs. 
The second term in the braces of Eq.~(\ref{eqn:gbarg})
vanishes in the limit of $\Omega^2\to 0+$. 
Thus, in this limit, $d_1(K,\Omega^2)$ of Eq.~(\ref{eqn:deeone}) becomes approximately $-1/2$ for $K\ll 1$, which 
can be seen in Fig.~\ref{fig:cald}.

\section{Some details of Sect.~\ref{sec:general} \label{app:procgen}}
As in Appendix A of Ref.~\onlinecite{sphere},
we rewrite 
Eq.~(\ref{eqn:wtaupsi}) as $u=1+s$ by defining
\begin{equation}
u\equiv w/\tau\quad {\rm and}\quad s\equiv C_2u^{1-2\beta}\tau^{-2\beta}\psi^2
\ ,\label{eqn:henkan}\end{equation}
where $C_2$ is defined below Eq.~(\ref{eqn:wtaupsi}).
Equation (\ref{eqn:f}) is rewritten as
\begin{equation}
f_{\rm R}(\psi)=
{k_BT_{\rm c} C_1\tau^{2\beta+\gamma}\over 2C_2\xi_0^2}u^{2\beta+\gamma-2}
\left( s+{s^2\over 6}\right)
\ .\label{eqn:fR}\end{equation}
Below, we consider 
$\psi^{(0)}(z)$ for $z>0$ and
regard $\xi$, $w$, $u$, and $s$ as being determined by $\psi^{(0)}(z)$. They become functions of $z$; 
$s(0+)$ denotes $s(z)$ in the limit of $z\to 0+$, for example.  
For simplicity of the following description in this appendix, we assume $h>0$ without loss of
generality.
On this assumption, $\psi^{(0)}(z)$ decreases from a positive value to zero as $z$ increases
from zero to $\infty$.
Mathematically, $\psi^{(0)}(z)$ calculated for $z>0$ can diverge to $\infty$ at a negative $z$ value, which we
write as $-l$. \\

With the aid of the scaling law $\gamma+\eta\nu=2\nu$,
we use Eqs.~(\ref{eqn:Cdef}) and (\ref{eqn:bulk}) to obtain
\begin{equation}
 {d\psi^{(0)}(z)\over dz}=
-{u^{\beta+\nu-1 }\tau^{\beta+\nu} \over 
\sqrt{C_2} \xi_0} \left( s+{s^2\over 6}\right)^{1/2}
\ .\label{eqn:dpdz}\end{equation}
Using
Eqs.~(\ref{eqn:Cdef}) and (\ref{eqn:dpdz}), we can rewrite Eq.~(\ref{eqn:bc}) as
\begin{equation}
{\Theta\over\tau^{\beta+\nu-\eta\nu}}=\left[1+s(0+)\right]^{\beta+\nu-\eta\nu-1}
\left[s(0+)+{s(0+)^2\over 6}\right]^{1/2}\label{eqn:bc2}
\ ,\end{equation}
where we use $\Theta\equiv \sqrt{C_2}\xi_0 h/\left( k_{\rm B} T_{\rm c} C_1\right)$. 
Substituting $u=1+s$ and Eq.~(\ref{eqn:dpdz}) into the derivative of 
the second entry of Eq.~(\ref{eqn:henkan}) with respect to $z$,
we obtain 
\begin{equation}
-\xi_\infty {d\sqrt{s}\over dz} = {\left(1+s\right)^{\nu+(1/2)}\left[s+\left(s^2/6\right)\right]^{1/2}\over 1+2\beta s}
\ ,\label{eqn:rootsdz}\end{equation}
which implies that $s(z)$ decreases monotonically as $z$ increases.
From this equation, we can obtain $dz/(ds)$, which is integrated 
from $s=\infty$ to $s(0+)$ to yield $l$.  \\

Using the material constants mentioned in Sect.~\ref{sec:genfree}, 
we solve Eq.~(\ref{eqn:bc2}) numerically and plot the results
with circles in Fig.~\ref{fig:pszeroh7}(a). Substituting these numerical results
into the second equation of Eq.~(\ref{eqn:henkan}) with the aid of
$u=1+s$, we obtain the results represented by the circles of Fig.~\ref{fig:pszeroh7}(b).
We find that $\psi^{(0)}(0+)$ reaches a plateau for sufficiently small $\tau$,
in contrast with the result calculated in terms of the Gaussian model (dashed curve).
Using the results of $s(0+)$, as mentioned below Eq.~(\ref{eqn:rootsdz}), we calculate $l$ 
numerically to plot the results with crosses in Fig.~\ref{fig:zeh7}.  
The first equation of Eq.~(\ref{eqn:henkan}) means
$w=\tau(1+s)$. Combining this with $w\equiv \xi_0^{1/\nu}\xi^{-1/\nu}$, we obtain
\begin{equation}
\xi(0+)=\xi_0 \left\{\tau\left[1+s(0+)\right]\right\}^{-\nu}
\ ,\label{eqn:xi0+}\end{equation}
which is plotted in Fig.~\ref{fig:zeh7}.
We find that each of $l$ and $\xi(0+)$ reaches a plateau for sufficiently small $\tau$, 
in contrast with $\xi_\infty$. \\

\begin{figure}
\includegraphics[width=8cm]{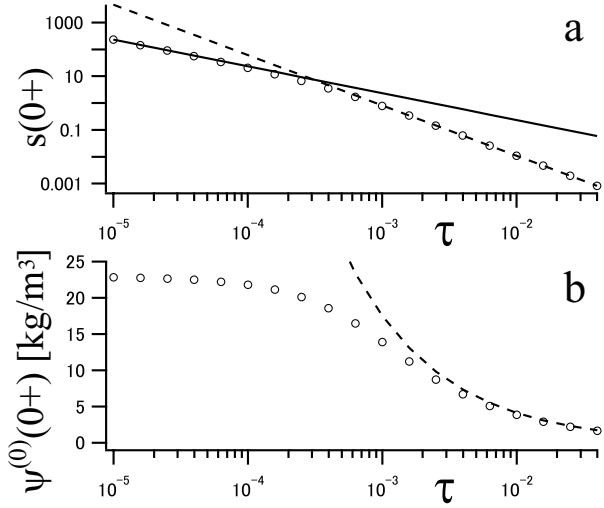}
\caption{\label{fig:pszeroh7} 
(a) Numerical results of $s(0+)$ plotted against $\tau$ with circles.
They are obtained from Eq.~(\ref{eqn:bc2}), and
agree with the dashed line and solid line
for $\tau\stackrel{>}{\sim}10^{-3}$ and $\tau\ll 10^{-3}$, respectively.
The former line represents
 $\Theta^2\tau^{-2(\beta+\nu-\nu\eta)}$, which comes from Eq.~(\ref{eqn:bc2})
for $s(0+)\ll 1$, while the latter represents Eq.~(\ref{eqn:s0+}).
(b) Numerical results of $\psi^{(0)}(0+)$ plotted against $\tau$ with circles.  The dashed curve
is calculated from Eq.~(\ref{eqn:phizero}), where $M$ and $\xi_{\rm c}$ are
respectively regarded as $k_{\rm B}T_{\rm c}C_1$ and $\xi_0 \tau^{-\nu}$.  
In (a) and (b), we use the material constants mentioned in Sect.~\ref{sec:genfree}. }
\end{figure}

Using $u=1+s$ and the second equation of Eq.~(\ref{eqn:henkan}), we have
\begin{equation}
\left(s\tau\right)^\beta\approx \sqrt{C_2}\psi^{(0)} \quad {\rm for}\ s\gg 1\ .
\label{eqn:stau}\end{equation}
Let us assume $s(0+)\gg 1$; we have
$s(z)\gg 1$ up to a positive value of $z$.  In this range of $z$, 
Eq.~(\ref{eqn:dpdz}) leads to ${\psi^{(0)}}'(z) \approx 
-\left(s\tau\right)^{\beta+\nu} / \left( \sqrt{6C_2}\xi_0\right)$ because $u\approx s$, 
and we obtain
\begin{equation}
\psi^{(0)}(z)\approx {1\over \sqrt{C_2} }\left[{\sqrt{6} \beta\xi_0\over \nu \left(z+l\right)}\right]^{\beta/\nu}
\label{eqn:jcppsi}\end{equation}
with the aid of Eq.~(\ref{eqn:stau}).  Equation (\ref{eqn:jcppsi})
is the same as Eq.~(2.15) of Ref.~\onlinecite{JCP} for $\tau=0$ and $\lim_{z\to \infty}\psi^{(0)}(z)=0$ because of
the statement below its Eq.~(3.15). \\

Suppose that $\tau$ is small enough to give
$l\ll \xi_\infty$.  
Judging from Eqs.~(\ref{eqn:stau}) and (\ref{eqn:jcppsi}), we have $s(\xi_\infty)\approx 1$, 
and have $s(z)\gg 1$ for $z\ll \xi_\infty$.  Thus, the range of
$l\ll z\ll \xi_\infty$ emerges for sufficiently small $\tau$, and in this range
Eq.~(\ref{eqn:jcppsi}) approximately yields
\begin{equation}
\psi^{(0)}(z)\propto \left( {\xi_0\over z}\right)^{\beta/\nu}  = \tau^\beta \left({z\over \xi_\infty}\right)^{-\beta/\nu}
\ ,\label{eqn:univ}\end{equation}
which is consistent with the universal form \cite{diehl97, diehl94, smdilan, smock}.
The origin appears to be shifted by distance $l$, which is negligible in the range of $l\ll z\ll \xi_\infty$.  
This shift and the range are mentioned in Refs.~\onlinecite{smdilan} and \onlinecite{rudjas}. 
\\

Because $s(z)$ decreases monotonically to zero as $z$ increases,
we have $s(z)\ll 1$ beyond a value of $z$ regardless of how small $\tau(>0)$ may be.  
In this range of $z$, noting that $u\approx 1$,
we substitute the second equation of Eq.~(\ref{eqn:henkan}) into Eq.~(\ref{eqn:dpdz}) and
obtain 
\begin{equation}
{d\psi^{(0)} \over dz}\approx -{\psi^{(0)}\over \xi_\infty} \quad, {\it i.e.\/},\ \psi^{(0)}(z)\propto e^{-z/\xi_\infty}
\ .\label{eqn:gaussian}\end{equation} 
Hence, 
the same exponential decay as that of Eq.~(\ref{eqn:phizero}) in the Gaussian model 
appears beyond some distance from the membrane, which is 
well known \cite{diehl94,smdilan} and is also mentioned in Appendix \ref{app:unpgen}.
When $\tau$ is large enough to give $s(0+)\ll 1$, Eq.~(\ref{eqn:gaussian}) 
holds from very near the membrane.  
Thus, for the material constants used here, 
the Gaussian model is valid for $\tau\stackrel{>}{\sim} 3 \times 10^{-3}$ judging from Fig.~\ref{fig:pszeroh7}(a) \cite{later}.
In this range of $\tau$, $\xi(0+)$ agrees with $\xi_\infty$ in Fig.~\ref{fig:zeh7}, while
$\psi^{(0)}(0+)$ agrees with the result of the Gaussian model in Fig.~\ref{fig:pszeroh7}(b). 
\\

Suppose that $\tau(>0)$ is small enough to give $s(0+)\gg 1$. 
Equations (\ref{eqn:xi0+})-(\ref{eqn:jcppsi}) give
\begin{equation}
l \approx {\sqrt{6}\beta \xi(0+)\over \nu}
\ ,\label{eqn:Lxi0}\end{equation}
which is approximately equal to $1.27 \xi(0+)$. This can be seen from Fig.~\ref{fig:zeh7}.  Thus, 
$l$ cannot be neglected totally, considering that
 the coarse-graining is performed 
up to $\xi(0+)$ near the membrane.
The rhs of Eq.~(\ref{eqn:bc2}) is approximated to 
$s(0+)^{\beta+\nu-\nu\eta}/\sqrt{6}$, which leads to
\begin{equation}
s(0+)\approx \left( \sqrt{6}\Theta\right)^{1/(\beta+\nu-\nu\eta)}\tau^{-1} 
\ . \label{eqn:s0+}\end{equation}
This is combined with Eq.~(\ref{eqn:xi0+}) to give $\xi(0+)\approx 
\xi_0\left( \sqrt{6}\Theta \right)^{\nu/(\nu\eta-\beta-\nu)}$.
We use Eq.~(\ref{eqn:Cdef}), $\epsilon=1$, and  $(\eta-3)\nu/(\nu\eta-\beta-\nu)=2$ to obtain
\begin{equation}
{\xi(0+)^3 \over M_{00}}\approx  {k_{\rm B}T_{\rm c}\over 4\pi^2 h^2}
\ .\label{eqn:tuika}\end{equation}

\bigskip
To obtain the circles in Fig.~\ref{fig:dC7h7bun}(a) and the symbols except for the solid circles
in Fig.~\ref{fig:dC7h7bun}(b), we use $u=1+s$ and Eq.~(\ref{eqn:fR}) in the numerical integration of Eq.~(\ref{eqn:Ddef2}), 
where we obtain $s(z)$ by solving
 Eqs.~(\ref{eqn:bc2}) and (\ref{eqn:rootsdz}) numerically for small $z$ and  
by utilizing Eq.~(\ref{eqn:gaussian}) for large $z$. 
Let us assume 
that $\tau(>0)$ is small enough to give $s(z)\gg 1$ 
up to a sufficiently large value of $z$.
Then, the contribution from
the region beyond this $z$ value is negligible
in the integral of Eq.~(\ref{eqn:Ddef2}).
Thus, we substitute 
Eq.~(\ref{eqn:fR}) with $s\gg 1$ into the integral to obtain  
\begin{equation}
D(k)\approx { k_{\rm B} T_{\rm c}k^2\over 9u^*\xi_0^3}
\int_0^\infty dz\ \left(s\tau\right)^{3\nu}
\label{eqn:Dapp}\end{equation}
by using $\epsilon=1$ and $2\beta+\gamma=3\nu$. 
Because of Eq.~(\ref{eqn:stau}), 
$\left(s\tau \right)^\beta$ is approximately equal to $\sqrt{C_2}$ multiplied by
the rhs of Eq.~(\ref{eqn:jcppsi}).
Thus, with the aid of Eq.~(\ref{eqn:Lxi0}), Eq.~(\ref{eqn:Dapp}) gives
\begin{equation}
D(k)\approx {\sqrt{6}\beta k_{\rm B}T_{\rm c} k^2  \over 2 \nu \pi^2 \xi(0+)^2} 
 \int_{1}^\infty dz\ z^{-3}\ ,\label{eqn:Dappapp}
\end{equation}
which yields Eq.~(\ref{eqn:Dapp2}).
Equations (\ref{eqn:Dapp2}), (\ref{eqn:xi0+}), and (\ref{eqn:s0+})
lead to $D(k)\propto h^{4/3}$ approximately.  
\\

\section{\label{sec:naive}  Naive average in the general model }
Some details leading to Eq.~(\ref{eqn:main3})
are shown here; 
$M_0$, $M_{0}'$, and $M''_{0}$ denote $M(\varphi^{(0)})$, $M'(\varphi^{(0)})$, and  
$M''(\varphi^{(0)})$, respectively. 
As argued for a mixture in contact with a flat wall \cite{Cahn},  Eqs.~(\ref{eqn:hatmudef})
and (\ref{eqn:phisurface}) respectively yield 
\begin{equation}
f'(\varphi^{(0)})-
M_0{\varphi^{(0)}}''-{M'_0\over 2}\left|{\varphi^{(0)}}'
\right|^2=\mu^{(0)}
\label{eqn:equalmu}
\end{equation}
and the boundary condition
$M_{0}{\varphi^{(0)}}' =\mp h$ at $z\to 0\pm$, at the order of $\varepsilon^0$.
These equations can also be obtained by minimizing Eq.~(\ref{eqn:glw}) around a planar membrane
with the difference in the total mass between the two components maintained \cite{ofk}. They
lead to Eqs.~(\ref{eqn:bulk}) and (\ref{eqn:bc}), respectively \cite{com2}, and yield Eq.~(\ref{eqn:phizero})
in the Gaussian model.  
At the order of $\varepsilon$, Eq.~(\ref{eqn:hatmudef}) gives
\begin{equation}
\left[ M_0\Delta +M_0'{\varphi^{(0)}}' {\partial\over\partial z} + N(z) \right]\varphi^{(1)} 
=-\mu^{(1)}
\ ,\label{eqn:mu1}\end{equation}
where $N(z)$ is defined as
$-f''(\varphi^{(0)})+M_0'{\varphi^{(0)}}''+
M_0''\left|{\varphi^{(0)}}'\right|^2/2$.
Equation (\ref{eqn:mu1}) gives Eq.~(\ref{eqn:newouter3})
in the Gaussian model.  The lhs of Eq.~(\ref{eqn:mu1})
vanishes when $\varphi^{(1)}$ is replaced by ${\varphi^{(0)}}'$.  This is verified by differentiating
Eq.~(\ref{eqn:equalmu}) with respect to $z$, and is used later.  \\

Defining ${\check \Omega}$ as Eq.~(\ref{eqn:glw}) in which
$M$ and $f$ are respectively given by Eqs.~(\ref{eqn:Cdef}) and (\ref{eqn:f}), we
suppose that $\psi$ and $\zeta$ deviate from the unperturbed profile $\psi^{(0)}$ and zero, respectively.
We write $\psi_1$ for the former deviation.  Below,
the resultant deviation of ${\check \Omega}$, denoted by $\delta{\check\Omega}$, is calculated
up to the second order with respect to $\psi_1$ and $\zeta$.
 The prime indicates the derivative,    
as stated below Eq.~(\ref{eqn:hatmudef}).  
Noting that Eq.~(\ref{eqn:equalmu}) with $\mu^{(0)}=0$ and its subsequent boundary condition hold, we find
that $\delta{\check\Omega}[\psi_1,\zeta]$ is given by
 Eq.~(A9) of  Ref.~\onlinecite{pre} with the integrand of its first integral
 and that of the second line being respectively replaced by \cite{a9}
\begin{equation}
\left( {f_{\rm R}''\over 2}  +{M_0''{\psi^{(0)}}'^2\over 4}\right)\psi_1^2 
+M_0'{\psi^{(0)}}' \psi_1{\partial \psi_1 \over\partial z}+{M_0 \left|\nabla\psi_1\right|^2\over 2} 
\end{equation}
and $W_{00} h\zeta^2 /M_{00} - W_{00} \zeta \left[\psi_1(\bm{x},\zeta+) - \psi_1(\bm{x},\zeta-)\right]$, 
where $W_{00}$ is defined as
$M_0'\left(h/M_0\right)^2 +M_0{\psi^{(0)}}''$ evaluated at $z=0+$.
This result of $\delta{\check\Omega}$ becomes Eq.~(A9) of  Ref.~\onlinecite{pre}
in the Gaussian model. 
We write $\delta{\check\Omega}^+$ for the sum of $\delta{\check\Omega}$ and the contribution from
the membrane energy, which is independent of $\psi_1$ and consists of the bending energy and
the energy involving $p_{\rm m}^{(0)}$.
In this static theory for the open system \cite{com2}, 
the probability density of $\psi_1$ and $\zeta$ is proportional to the Boltzmann weight,
$\exp{\left[ -\delta{\check \Omega}^+/(k_{\rm B}T)\right] }$.  Integrating this density with respect to
$\psi_1$ yields the probability density of $\zeta$, from which we can find its variance, 
{\it i.e.\/}, the MSA.  
As in the appendix of Ref.~\onlinecite{pre}, we instead obtain this probability density by taking
the minimum of $\delta{\check\Omega}[\psi_1,\zeta]$ for a given $\zeta$.  
This approximation is valid when we assume  $k\xi_\infty\ll 1$
to neglect the critical fluctuation of the mixture, which is large at smaller length scales.  The stationary condition 
is given by Eq.~(\ref{eqn:mu1}) with its rhs put equal to zero and Eq.~(\ref{eqn:bcphi}) if
$\varphi^{(1)}$ and $\zeta^{(1)}$ are respectively replaced by $\psi_1$ and $\zeta$. 
Let us write $\psi^*_1$ for the solution of the stationary condition,
which enables us to calculate $\delta{\check\Omega}[\psi_1^*,\zeta]$ as
\begin{eqnarray}&&
l_{\rm p}^2\sum_{\bm{k}} \left\{ k^2f_{\rm s}(\varphi^{(0)}(0+) {\hat\zeta}(-\bm{k})
{\hat\zeta}(\bm{k}) \right.\nonumber\\&&\quad 
\left. +  W_{00} {\hat\zeta}(-\bm{k})
\left[{h {\hat\zeta}(\bm{k}) \over M_{00}}
-{\hat \psi}_1^*(\bm{k},0+)
\right]\right\}\ .
\label{eqn:deltaOmegamin}\end{eqnarray}
This gives the minimum of $\delta{\check\Omega}$ for a given $\zeta$.\\

Considering the statement below Eq.~(\ref{eqn:mu1}), we find
that ${\hat \psi}^*_1(\bm{k})$ tends to $-{\hat \zeta}(\bm{k}) {\psi^{(0)}}'(z)$ in the limit of $k\to 0+$.
Let us define $\chi(\bm{k},z)$ as the quotient of ${\hat \psi}_1^*(\bm{k},z)$ divided by $-{\hat \zeta}(\bm{k}) {\psi^{(0)}}'(z)$;
$\chi(\bm{k},z)$ tends to unity as $k$ approaches zero.  The stationary condition is rewritten as\cite{gaussapp} 
\begin{equation}
\chi''+ \left({d\over dz} \ln{ M_0{{\psi^{(0)}}'}^2} \right) \chi' =k^2\chi \quad {\rm for}\ z\ne 0
\label{eqn:chieq}\end{equation} 
and $W_{00}(\chi-1)=h\chi'$ at $\ z\to 0\pm$, where
the prime indicates differentiation with respect to $z$.
Noting that ${\hat \psi}_1^*(\bm{k},z)$ is odd with respect to $z$,
we below consider $\chi(\bm{k},z)$ for $z> 0$.    
Because of the statement above Eq.~(\ref{eqn:chieq}), we can
approximate the rhs of Eq.~(\ref{eqn:chieq}) to $k^2$
for a sufficiently small $k$.  This is realized by expanding $\chi$ with respect to $k$.
Within this approximation, we obtain
\begin{equation}
\chi'(\bm{k},z)=
{ -k^2 \int_z^\infty dz'\  
M(\varphi^{(0)}(z')) {\psi^{(0)}}'(z')^2 \over M(\varphi^{(0)}(z)) {\psi^{(0)}}'(z)^2}
\ ,\label{eqn:chiprime}\end{equation}
where we use the condition that ${\hat \psi}_1^*$, and thus $\chi{\psi^{(0)}}'$, vanish in the limit of $z\to \infty$.
Noting the boundary condition mentioned below Eq.~(\ref{eqn:chieq}), we thus arrive at
\begin{eqnarray}&&
{\hat \psi}_1^*(\bm{k},0+)-{h {\hat \zeta}(\bm{k})\over M_{00}}\nonumber\\ &&=
-{k^2{\hat\zeta}(\bm{k}) \over W_{00}}
\int_0^\infty dz\ M(\varphi^{(0)}(z))\left( {d \varphi^{(0)}\over dz }\right)^2 \ ,\label{eqn:psi1sol}
\end{eqnarray}
where we use Eq.~(\ref{eqn:bc}).
We can substitute Eq.~(\ref{eqn:psi1sol}) into
Eq.~(\ref{eqn:deltaOmegamin}) for sufficiently small $k$.  
Adding the terms of the membrane energy, we 
find the probability density of $\hat{\zeta}(\bm{k})$.
For a tensionless membrane, 
the term involving $f_{\rm s}$ in Eq.~(\ref{eqn:deltaOmegamin}) 
cancels with the term involving $p_{\rm m}^{(0)}$ because of Eq.~(\ref{eqn:lat}),
and the MSA for sufficiently small $k$ is found to be given by
Eq.~(\ref{eqn:main3}) together with Eq.~(\ref{eqn:Ddef1}).


\section{Supplement on hydrodynamic effects in the general model \label{app:hydgen}}
In Ref.~\onlinecite{pre}, we study the MSA in the Gaussian model
by assuming sufficiently weak preferential attraction.  
On this assumption, as shown below, we can use the general model to consider the 
hydrodynamic effects of the ambient mixture to derive
Eq.~(\ref{eqn:main3}) with $D$ being replaced by $2D$.  
This result almost coincides with Eq.~(\ref{eqn:mainato})
with $\xi_{\rm c}$ being replaced by $\xi(0+)$, considering the statement below
Eq.~(\ref{eqn:Dapp2ato}).
In the Gaussian model, as discussed in Sect.~\ref{sec:gaussianthree},
the condition under which the hydrodynamic effects of the mixture are negligible
is given by  $k^3\ll 4h^2/(c_{\rm b}M)$.  The discussion below Eq.~(\ref{eqn:kikamoto2}) 
suggests that this condition should remain valid 
beyond the regime of the Gaussian model.
Judging from these results, the reduction in the restoring force by approximately half, mentioned below Eq.~(\ref{eqn:mainhalf}),
is expected to continue beyond the regime of the Gaussian model.  \\

In general, we define $\Omega \equiv \omega \sqrt{\rho \xi(0+)^{3}}/ 
\left(\lambda {k\sqrt{c_{\rm b}}}\right)$, where $\lambda$ is defined as
\begin{equation}
\lambda\equiv {\sqrt{M_{00}} \xi(0+)^{5/2} \over \sqrt{c_{\rm b}}} {\varphi^{(0)}}''(0+)\ .
\label{eqn:lamdef}\end{equation}
With the aid of Eq.~(\ref{eqn:phizero}), we find that these definitions 
give Eq.~(\ref{eqn:Lamgauss})
in the regime of the Gaussian model, where 
$\xi(0+)$ and $M_{00}$ can be identified respectively with $\xi_{\rm c}$ and  
$k_{\rm B}T_{\rm c}C_1$.  
When $h$ does not vanish and $s(0+)$ is much larger than unity, 
we can write ${\psi^{(0)}}''(0+)$ in terms of ${\psi^{(0)}}'(0+)$ by using Eq.~(\ref{eqn:jcppsi}).
Applying this result in Eq.~(\ref{eqn:lamdef}), we then derive 
\begin{equation}
\lambda \approx {h\xi(0+)^{3/2} \over \sqrt{6c_{\rm b}M_{00}}}\left(1+{\nu\over\beta}\right) 
\label{eqn:preineq}\end{equation}
with the aid of Eqs.~(\ref{eqn:bc}) and (\ref{eqn:Lxi0}).
Through the same discussion given below Eq.~(\ref{eqn:Dapp2ato}),
we find that the rhs of Eq.~(\ref{eqn:preineq}) approximately equals that of the first equation of 
Eq.~(\ref{eqn:Lamgauss}) with $\xi_{\rm c}$ being replaced by $\xi(0+)$.  
Using Eq.~(\ref{eqn:tuika}), we find that Eq.~(\ref{eqn:preineq}) gives
\begin{equation}
\lambda^2\approx {k_{\rm B}T_{\rm c}\over 24\pi^2 c_{\rm b}} \left(1+{\nu\over \beta}\right)^2
\label{eqn:lamlam}\end{equation}
for sufficiently small $\tau$.  
Thus, for $h\ne 0$, $\lambda$ in the limit of $\tau\to 0+$ is independent of $h$.  \\

Below, we generalize the definitions of $Z$ and $K$ in  
Sect.~\ref{sec:gauss} to use $Z\equiv z/\xi(0+)$ and $K\equiv k\xi(0+)$;
they respectively become $Z\equiv z/\xi_{\rm c}$ and $K=k\xi_{\rm c}$ in the Gaussian model.
We also generalize $G$, $Q$ and $\Xi$ as
\begin{equation}
G\left(\bm{k}, Z, \omega\right)\equiv \displaystyle{{{\tilde \varphi}^{(1)}\left(\bm{k},z,\omega\right)
\over \xi(0+) {\varphi^{(0)}}''(0+) {\tilde \zeta}^{(1)}\left(\bm{k},\omega\right)}}
\ ,\label{eqn:defG2}\end{equation}
\begin{equation}
Q\left(\bm{k}, Z, \omega\right)\equiv \displaystyle{{\xi(0+){\tilde \mu}^{(1)}\left(\bm{k},z,\omega\right)
\over M_{00}{\varphi^{(0)}}''(0+) {\tilde \zeta}^{(1)}\left(\bm{k},\omega\right)}}
\ ,\label{eqn:defQ2}\end{equation}
and \begin{equation}
\Xi(Z)\equiv {{\varphi^{(0)}}'(z)\over  \xi(0+) {\varphi^{(0)}}''(0+)}\ .\label{eqn:Xidef}\end{equation} 
The corresponding equations in the Gaussian model are found in Eqs.~(\ref{eqn:defG}) and (\ref{eqn:Xigauss}), 
where Eq.~(\ref{eqn:phizero}) is available.
Equations (\ref{eqn:pZKU})--(\ref{eqn:partUU})
remain valid after these redefinitions.  We can derive Eq.~(\ref{eqn:partGout})
by using Eq.~(\ref{eqn:mu1}) instead of Eq.~(\ref{eqn:newouter3}).  With the aid of
Eq.~(\ref{eqn:mu1}) and the boundary condition mentioned below Eq.~(\ref{eqn:equalmu}), Eq.~(\ref{eqn:phiz0})
is found to be valid if $M$ on its rhs is replaced by $M_{00}$;
${\tilde \varphi}^{(1)}$ equals $-h{\tilde \zeta}^{(1)}/M_{00}$ in the limit of
$z\to 0-$ in addition.   Thus, we can derive Eq.~(\ref{eqn:bcphi}) 
in the limit of $L\to 0+$ even when $M$ depends on $\varphi$.
We use Eq.~(\ref{eqn:phiz0}) with the replacement above
and Eq.~(\ref{eqn:equalmu}) to derive  Eq.~(\ref{eqn:tilFz}) in the general model, where
Eq.~(10) of Ref.~\onlinecite{pre} remains valid.  
\\

Replacing  ${\varphi}^{(1)}$ and ${\mu}^{(1)}$ by their respective outer solutions 
in Eq.~(\ref{eqn:mu1}), we obtain
\begin{equation}
Q_{\rm out}={M_0\Xi\over M_{00}}\left(\partial_Z^2-K^2\right) U_{\rm out}
 +{\partial_Z U_{\rm out}\over M_{00}\Xi} {d M_0\Xi^2 \over dZ} 
\label{eqn:mu1karanew}\end{equation}
from Eq.~(\ref{eqn:newouter}) with the aid of
the fact mentioned below Eq.~(\ref{eqn:mu1}).   The equation that
Eq.~(\ref{eqn:mu1karanew}) becomes in the Gaussian model can be obtained from
Eqs.~(\ref{eqn:newouter}), (\ref{eqn:phizero}), and (\ref{eqn:newouter3}).  
In general, from Eq.~(\ref{eqn:Usol1}), we have 
$U_{\rm out}(Z)=U(Z)=e^{-KZ}$ for $Z>0$ up to the order of $1/\Omega$. 
Substituting this into Eq.~(\ref{eqn:mu1karanew}) yields
$Q_{\rm out}$ up to the order of $1/\Omega$.  
Substituting this approximate result 
into Eq.~(\ref{eqn:partZUout}) yields 
\begin{equation}
\lim_{Z\to 0+}\partial_Z U_{\rm out}\approx -K-{\Xi(0+)^2 K\over \Omega^2} +{k^2 \xi(0+)\over \rho\omega^2} D(k)
\ .\label{eqn:preKAdg}\end{equation}  
We substitute Eq.~(\ref{eqn:preKAdg}) into Eq.~(\ref{eqn:partUU}), substitute
the result into Eq.~(\ref{eqn:tilFz}), and then use Eq.~(\ref{eqn:maku}) to find that
the MSA is given by  Eq.~(\ref{eqn:main3}) with $D$ replaced by $2D$.  
Exactly speaking, 
$D(k)$ here is given by Eq.~(\ref{eqn:Ddef1}) 
with the integrand being multiplied by 
$e^{-2kz}$, which is negligible for $K\ll 1$ however.  \\

From Eqs.~(\ref{eqn:pZKU}) and (\ref{eqn:mu1karanew}), we obtain
\begin{eqnarray}&&
\left(1-{M_{00} \Omega^2 \over M_0\Xi^2} \right)(\partial_Z^2-K^2)U_{\rm out}
\nonumber\\ &&\quad
=-\left[  {M_0'{\psi^{(0)}}'\xi(0+)\over M_0}  +{2\Xi'\over\Xi}\right]   
\partial_Z U_{\rm out}\ ,
\label{eqn:kikamoto2}\end{eqnarray}
where $\psi^{(0)}$ denotes the solution of Eqs.~(\ref{eqn:bc2}) and (\ref{eqn:rootsdz}).
The boundary conditions at $Z\to 0+$ and $Z\to \infty$
remain the same as stated below Eq.~(\ref{eqn:kikamoto}).
Being proportional to $f_{\rm R}(\psi^{(0)}(z))$ 
because of Eq.~(\ref{eqn:bulk}), $M_0\Xi^2$ 
decreases to zero as $z$ increases to $\infty$. 
Thus, when $\Omega^2$ is smaller than $\Xi(0+)^2$,
the difference in the first parentheses on the lhs of Eq.~(\ref{eqn:kikamoto2}) vanishes
at the value of $Z$ making $M_0 \Xi^2/\Omega^2$ equal to $M_{00}$. \\

Equation (\ref{eqn:kikamoto2}) is reduced to Eq.~(\ref{eqn:kikamoto}) in the Gaussian model.
The difference in the first parentheses on the lhs of Eq.~(\ref{eqn:kikamoto})
is approximately equal to $-2\left(Z-Z^*\right)$ around $Z=Z^*$, by which we can predict
logarithmic divergence at $Z=Z^*$ in $U_{\rm out}$ in the Gaussian model.
Applying a similar discussion to Eq.~(\ref{eqn:kikamoto2}), we can predict logarithmic divergence at the value of
$Z$ making $M_0 \Xi^2/\Omega^2$ equal to $M_{00}$.
We introduce $d_1$ at the second term in the braces of Eq.~(\ref{eqn:perpendrep}) in the Gaussian model.  
The definition of $d_1$ can be generalized so that
this term is given by $2D(k)\xi(0+)^4 \left(1+d_1\right) /c_{\rm b}$.  
\\

At the end of Appendix \ref{app:stress}, we show $d_1(K,0+)\approx -1/2$ for small $K$ by using
the solution of Eq.~(\ref{eqn:kikamoto}) in the Gaussian model.
A simpler but approximate discussion to this result is as follows.
Suppose $\Omega^{2}\ll 1$;  Eq.~(\ref{eqn:kikamoto}) is approximated to be
$(\partial_Z^2-K^2)U_{\rm out}
 = 2\partial_Z U_{\rm out}$ 
up to a sufficiently large value of $Z$.  
This yields $U_{\rm out}(Z) \approx e^{(1-K_1)Z}$ with the aid of the boundary conditions of $U_{\rm out}$. 
This approximate result can be alternatively obtained as discussed at the end of Sect.~\ref{sec:gaussianproc}.
Using it in Eqs.~(\ref{eqn:partUU}) and (\ref{eqn:d1tochu}),
we obtain $d_1(K,0+)\approx -1/2$ for small $K$.  \\

Let us consider Eq.~(\ref{eqn:kikamoto2}) similarly to discuss $d_1$ 
for sufficiently small $\tau$.
We here neglect the first term in the braces, 
considering the weak dependence of $M$ on $\varphi$.
Using Eqs.~(\ref{eqn:jcppsi}) and (\ref{eqn:Xidef}), we can calculate $\Xi'/\Xi$ approximately.
We introduce $Z_0\equiv l/\xi(0+)$, which approximately equals $\sqrt{6}\beta/\nu$ 
for small $\tau$ because of Eq.~(\ref{eqn:Lxi0}).  With the aid of 
$2\beta/\nu=1+\eta$,
Eq.~(\ref{eqn:kikamoto2}) is thus
approximated to 
$\left(\partial_Z^2-K^2\right)U_{\rm out} = \left(3+\eta\right)\partial_Z U_{\rm out} / \left( Z+Z_0\right)$
up to a sufficiently large value of $Z$. In this range of $Z$, $U_{\rm out}(Z)$ is found to be proportional approximately to the product of
$\left(Z+Z_0\right)^{2+(\eta/2)}$ and  ${\cal K}_{2+(\eta/2)} \left(K(Z+Z_0)\right)$, 
where ${\cal K}_q$ represents the modified Bessel function. 
By using  $U_{\rm out}(0+)=1$, 
we find \begin{equation}
\lim_{Z\to 0+} \partial_Z U_{\rm out}= -K {{\cal K}_{1+(\eta/2)}(KZ_0)\over {\cal K}_{2+(\eta/2)}(KZ_0)}
\ ,\end{equation}
which tends to $-K^2 Z_0/(2+\eta)$ as $KZ_0$ approaches zero. 
From this, we use Eq.~(\ref{eqn:Dapp2}) to
obtain $d_1(K,0+)\approx (-\eta-1)/(\eta+2) \approx -1/2$ for small $KZ_0=kl$.  \\

From these discussions, we can expect that $d_1$ has a finite peak at $\Omega^2=\Xi(0+)^2$ when plotted against 
$\Omega^2$.  We use Eqs.~(\ref{eqn:jcppsi}),  (\ref{eqn:Lxi0}), and (\ref{eqn:Xidef}) to find
$\Xi(0+)^2 \approx 6\beta^2 / (\beta+\nu)^2  \approx 0.7 $
when $\tau$ is small enough to give $s(0+)\gg 1$.    
Then, the peak position 
would not be so much shifted from the one in the Gaussian model, $\Omega^2=1$. 
Thus, Fig.~\ref{fig:cald} is expected to remain almost unchanged,
if $\xi_{\rm c}$ is replaced by $\xi(0+)$, beyond the regime of the Gaussian model.

\begin{figure*}
\includegraphics[width=14cm]{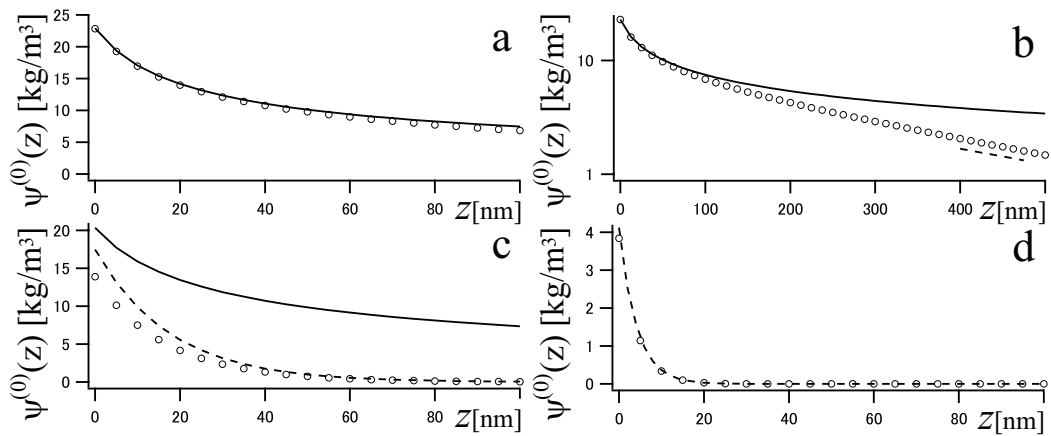}
\caption{\label{fig:psh7} 
Numerical results of the unperturbed profile $\psi^{(0)}(z)$ (circles).
The solid curve represents Eq.~(\ref{eqn:jcppsi}).  
The material constants mentioned in Sect.~\ref{sec:genfree} are used.
We use $\tau=10^{-5}$ to have
$\xi_\infty=3.14\times 10^{2}$ nm  in (a) and (b).
The linear plot is given in (a), while
the semi-logarithmic plot for a wider range of $z$ is given in (b), where the dashed line represents the slope
of $-1/\xi_\infty$.  We use $\tau=10^{-3}$ in (c)
and $10^{-2}$ in (d), which respectively give $\xi_\infty=1.75\times 10$ nm 
and $\xi_\infty=4.13$ nm.
The dashed curves in (c) and (d) represent Eq.~(\ref{eqn:phizero})
with $M$ and $\xi_{\rm c}$ being
respectively replaced by $k_{\rm B}T_{\rm c}C_1$ and $\xi_0 \tau^{-\nu}$. }
\end{figure*}

\section{Supplement on the unperturbed profile in the general model \label{app:unpgen}}
Solving Eqs.~(\ref{eqn:bc2}) and (\ref{eqn:rootsdz}) numerically,
we use Eq.~(\ref{eqn:henkan}) to calculate $\psi^{(0)}(z)$.  
In Figs.~\ref{fig:psh7}(a) and (b), where $\tau$ is sufficiently small,
we find that Eq.~(\ref{eqn:jcppsi}) gives a good approximation for small $z$; 
a characteristic length is given by $l$ and thus by $\xi(0+)$ because of Eq.~(\ref{eqn:Lxi0}).  
For large values of $z$ in Fig.~\ref{fig:psh7}(b), the numerical results agree with Eq.~(\ref{eqn:gaussian}).
Figures \ref{fig:psh7}(c) and (d) show the results for larger values of $\tau$. 
In the former, neither Eq.~(\ref{eqn:phizero}) nor Eq.~(\ref{eqn:jcppsi}) 
gives a good approximation. In the latter, $\tau$ is sufficiently large to validate the Gaussian model; 
Eq.~(\ref{eqn:phizero}) gives a good approximation.
For these large values of $\tau$, Eq.~(\ref{eqn:jcppsi}) cannot explain the numerical results for $z>0$ and
$l$ becomes meaningless to them.  In Fig.~\ref{fig:psh7}, as expected, 
the spatial region where  Eq.~(\ref{eqn:gaussian}) is valid is detached from the membrane when $\tau$ is
small enough to give $s(0+)\gg 1$.  Then, the difference in the distance from the critical point becomes significant
between the mixture parts near and far from the membrane.  Thus,
$\xi(0+)$ deviates from $\xi_\infty$ for small values of $\tau$ in Fig.~\ref{fig:zeh7}.
\\

Though data are not shown, when we use $h=10^{-6} \ {\rm m}^3/{\rm s}^2$
instead of $10^{-7} \ {\rm m}^3/{\rm s}^2$ in Figs.~\ref{fig:zeh7} and \ref{fig:pszeroh7}, 
Eq.~(\ref{eqn:s0+}) works well up to larger $\tau$ 
and the plateau regions of $\xi(0+)$, $l$, and $\psi^{(0)}(0+)$ extend up to 
larger $\tau$.  Then,
the plateau value of $\psi^{(0)}(0+)$ is larger, the plateau values of 
$l$ and $\xi(0+)$ are smaller, and the Gaussian model
ceases to be valid $\tau\stackrel{<}{\sim} 10^{-2}$, where $\xi_\infty$ is larger than approximately
$2$ nm.  These behaviors of $\psi^{(0)}(0+)$ and $\xi(0+)$ imply that, when $h$ is larger for sufficiently small $\tau$, the
mixture part near the membrane becomes more different from the part far from the membrane
in the distance from the critical point.  The difference cannot be described by the Gaussian model.


\begin{thebibliography}{99}
\bibitem{comBL} We can check this by deriving
a differential equation on the real part of $Q$ from Eq.~(59)
of Ref.~\onlinecite{pre}, using an inner variable, and neglecting small terms
inside the boundary layer.
\bibitem{com1} Equation (\ref{eqn:tilFz}) 
is also used in Ref.~\onlinecite{pre}.  On the rhs of Eq.~(33) in Ref.~\onlinecite{pre}, 
the last term in the braces cancels with the second term, while the first term  
gives Eq.~(67) of Ref.~\onlinecite{pre}.
\bibitem{later} For $\tau\stackrel{>}{\sim}3 \times 10^{-3}$, 
because $s(0+)\ll 1$ in Fig.~\ref{fig:pszeroh7}(a), the second term in the parentheses becomes
negligible in Eq.~(\ref{eqn:fR}), considering  the statement below Eq.~(\ref{eqn:rootsdz}).
\bibitem{enhan} Assuming $f_{\rm s}$ to be a linear function should amount to assuming
the bare value of the surface enhancement to be positive, as mentioned in Appendix A of Ref.~\onlinecite{sphere}.
\bibitem{a9} The numerator of the first term of the second line should be $h^2\zeta^2$ in 
Eq.~(A9) of Ref.~\onlinecite{pre}, whose $\delta \Omega$ and $l$ correspond respectively with $\delta {\check\Omega}$
and $l_{\rm p}$ here.
\bibitem{gaussapp} In the Gaussian model, 
Eq.~(\ref{eqn:chieq}) with its subsequent boundary condition is solved exactly to yield
$\chi(\bm{k},z)=e^{(1-K_1)Z}/K_1$, which leads to
Eq.~(A11) of Ref.~\onlinecite{pre}.
\end{thebibliography}
\end{document}